\documentstyle[prb,aps,epsf,eqsecnum]{revtex}
\begin{document}


\title{Vortex Physics in Confined Geometries}

\author{M. Cristina Marchetti \footnote{Corresponding author. Permanent address: Physics Department, 
Syracuse University, Syracuse, NY 13244; Tel. (315)443-2581; Fax (315)443-9103; 
email:mcm@physics.syr.edu}  and David R. Nelson}
\address{Lyman Laboratory of Physics, Harvard University, Cambridge, MA 01238}

\maketitle

\begin{center}
(\today)
\end{center}

\begin{abstract}

\vspace*{0.1in}

Patterned irradiation of cuprate superconductors with columnar defects allows a new 
generation of experiments which can probe the properties of vortex liquids by forcing them to 
flow in confined geometries. 
Such experiments can be used to distinguish experimentally between continuous 
disorder-driven glass transitions of vortex matter, such as the vortex glass or the
Bose glass transition, and nonequilibrium polymer-like glass transitions driven by
interaction and entanglement. 
For continuous glass transitions,
an analysis of such experiments that combines an inhomogeneous
scaling theory with the hydrodynamic description of viscous flow of vortex liquids
can be used to infer the critical behavior.
After generalizing vortex hydrodynamics to incorporate currents and field gradients
both longitudinal and transverse to the applied field, the critical exponents for all six vortex liquid
viscosities are obtained. In particular,
the shear viscosity is predicted to diverge as $|T-T_{BG}|^{-\nu z}$ at the Bose glass
transition, with $\nu\simeq 1$ and $z\simeq 4.6$ the dynamical critical exponent. The scaling behavior 
of the ac resistivity is also derived. As concrete examples of flux flow in confined geometries,
flow in a channel and in the Corbino disk geometry are discussed in detail. 
Finally, the implications of scaling for the hydrodynamic 
description of transport in the dc flux transformer
geometry are discussed.

\end{abstract}


\section{Introduction}
\label{sec:intro}

In the mixed state of cuprate superconductors the magnetic field is concentrated in an array of 
flexible flux bundles that, much like ordinary matter, can form crystalline, liquid
and glassy phases.\cite{crabtree_nelson,blatter,crabtree_NATO}  The dynamics of the flux-line array determines the resistive 
properties of the material and has therefore been the focus of much theoretical and experimental work. Of particular interest are the phase transitions connecting various forms of vortex matter.
For example, a first order melting transition is now believed to connect the crystalline Abrikosov 
flux lattice to a melted flux liquid. \cite{crabtree_nelson} 
If the barriers to line crossings are sufficiently high, rapidly cooled vortex
liquids can bypass the solid phase altogether and form an entangled polymer-like glass
phase. \cite{nelsonprl} The transition to such a polymer glass would occur when the system
gets trapped in a metastable state, much like the transition in ordinary window glass.

Other types of glasses are also possible because of pinning in 
disordered samples. It was suggested some time ago  that point disorder 
may drive a continuous transition from a vortex liquid to an isotropic vortex glass state,
with vanishing linear resistivity.\cite{mpaf,ffh} 
The existence of this transition is still in doubt, as
many of the early experiments claiming to observe it \cite{koch,gammel} were in fact dominated 
by twin boundary pinning. The vortex glass remains, however, a credible candidate for
the description of the disordered solid phase observed in untwinned crystals at high fields. \cite{lopez_krusin}

Correlated disorder, that is disorder that can pin vortex lines coherently along a
specific direction, is also very important in many materials. It can be created 
artificially via the introduction of columnar damage tracks created by heavy ion irradiation {--} a procedure that was shown to result in dramatic improvement in the pinning 
of vortex lines. \cite{budhani,koncz,civale} It can also be present in the material 
in the form of families of parallel twin planes that pass completely through the sample. \cite{hawley}
In samples with columnar disorder, if the concentration of damage tracks (assumed to pass 
completely through the sample) exceeds the number
of flux lines, there is a low-temperature ``Bose glass'' phase, in which every vortex is trapped on a
columnar defect  and the linear resistivity vanishes.\cite{drnvv}
At high temperatures the vortices delocalize in an entangled 
flux-line liquid. The high temperature liquid transforms into a Bose glass via a second order 
phase transition at $T_{BG}$. Samples with single families of twin boundaries also exhibit 
continuous transitions to anisotropic glass phases. \cite{crabtree,bariloche}

The Bose glass transition is the one that is best understood and experimentally characterized.
For this reason we will use it in much of the following as our prototype of a disorder-driven
continuous glass transition in vortex matter.
It has been studied theoretically by viewing the vortex line trajectories as the world
lines of two-dimensional quantum mechanical particles \cite{nelsonprl,drnvv}. The thickness of the superconducting
sample corresponds to the inverse temperature of the quantum particles. In thick samples the physics
of vortex lines pinned by columnar defects becomes equivalent to the low temperature properties
of two-dimensional bosons with point disorder. In the low temperature phase Bose glass phase
the vortices behave like localized bosons.  The entangled flux liquid phase is resistive and corresponds to a boson superfluid.
For simplicity, we confine our attention here to fields less that the ``matching field'' 
$B_\phi=n_{\rm pin}\phi_0$,
with $n_{\rm pin}$ the areal density of columnar pins and $\phi_0=\hbar c/2e$ the flux quantum.

Although an exact theory of such continuous transitions from the Bose glass or from the vortex glass
to the entangled flux liquid 
phase is not available, 
near the transition most physical properties can be described via a scaling theory in terms of just 
two undetermined critical exponents \cite{ffh,drnvv,drnleo,fwgf,lidmar,nverratum}.
The low temperature  disorder-dominated glass phase 
is characterized by a a correlation volume whose size diverges at the transition.
Anisotropic disorder, such as columnar defects or twin planes, results in an anisotropic
correlation volume. 
The most complicated  case is that of planar disorder, embodied for instance by a single family 
of parallel twin planes. In this case there are three correlation lengths that diverge at the
continuous glass transition temperature, $T_G$ (see Fig.~1).
The growth of correlations in the direction perpendicular to both the external field
and the twin planes is described by the correlation length
\begin{equation}
\label{xiperp} 
\xi_\perp(T)\sim|T-T_{G}|^{-\nu}.
\end{equation}
A second diverging length, $\tilde{\xi}_\perp(T)$, describes the extent of correlations in the direction perpendicular
to the external field, but parallel to the plane of the twin boundaries, 
\begin{equation}
\label{xiperp_tilde} 
\tilde{\xi}_\perp(T)\sim|T-T_{G}|^{-\tilde{\nu}}.
\end{equation}
Finally, there is a diverging correlation length along the external field
(here the $z$ direction), which is aligned with
the twin planes,
\begin{equation}
\label{xipar} 
\xi_\parallel(T)\sim\xi_\perp^\zeta\sim|T-T_{G}|^{-\nu\zeta},
\end{equation}
with $\zeta$ the anisotropy exponent. For planar disorder $\zeta=1+\tilde{\nu}/\nu$
(this constraint follows from the finiteness of $c_{11}$ at the transition). 
In materials with columnar pins aligned with the external field the correlation
volume is isotropic in the plane normal to the columns. Thus $\xi_\perp=\tilde{\xi}_\perp$
and there are two  correlation lengths that diverge at the Bose glass transition
transition temperature, $T_{BG}$, characterizing correlations  in the directions 
perpendicular ($\xi_\perp$) and parallel ($\xi_\parallel$) to the linear defects. 
In this case the anisotropy exponent
$\zeta$ has the value $\zeta=2$.
The same results are expected to hold for a {\it mosaic} of twin boundary planes,
all containing the field direction, on scales large compared to the mosaic size.
Finally, for the case of point disorder controlling the transition to the
vortex glass phase at $T_{VG}$, the correlation volume scaling is 
assumed to be 
isotropic.\cite{ffh} There is only a single diverging correlation length, $\xi_\perp$,
with $\zeta=1$, although it is hard to rule out different length scales parallel and 
perpendicular to the field direction.
The correlation time controlling the relaxation of critical 
fluctuations is assumed to diverge as 
\begin{equation}
\label{tau} 
\tau\sim\xi_\perp^z\sim |T-T_{G}|^{-z\nu},
\end{equation}
with $z$ the dynamical critical exponent. The values of the critical exponents $\nu$ and $z$
of course depend on the type of disorder and differ for instance for the Bose glass
and the vortex glass. 
For the Bose glass transition the critical exponents have been determined via simulations 
to be 
$\nu\simeq 1$ and $z\simeq 4.6$ \cite{wallin}.
%
%
\begin{center}
\vspace{0.2in}
\epsfxsize=3.5in
\epsfbox{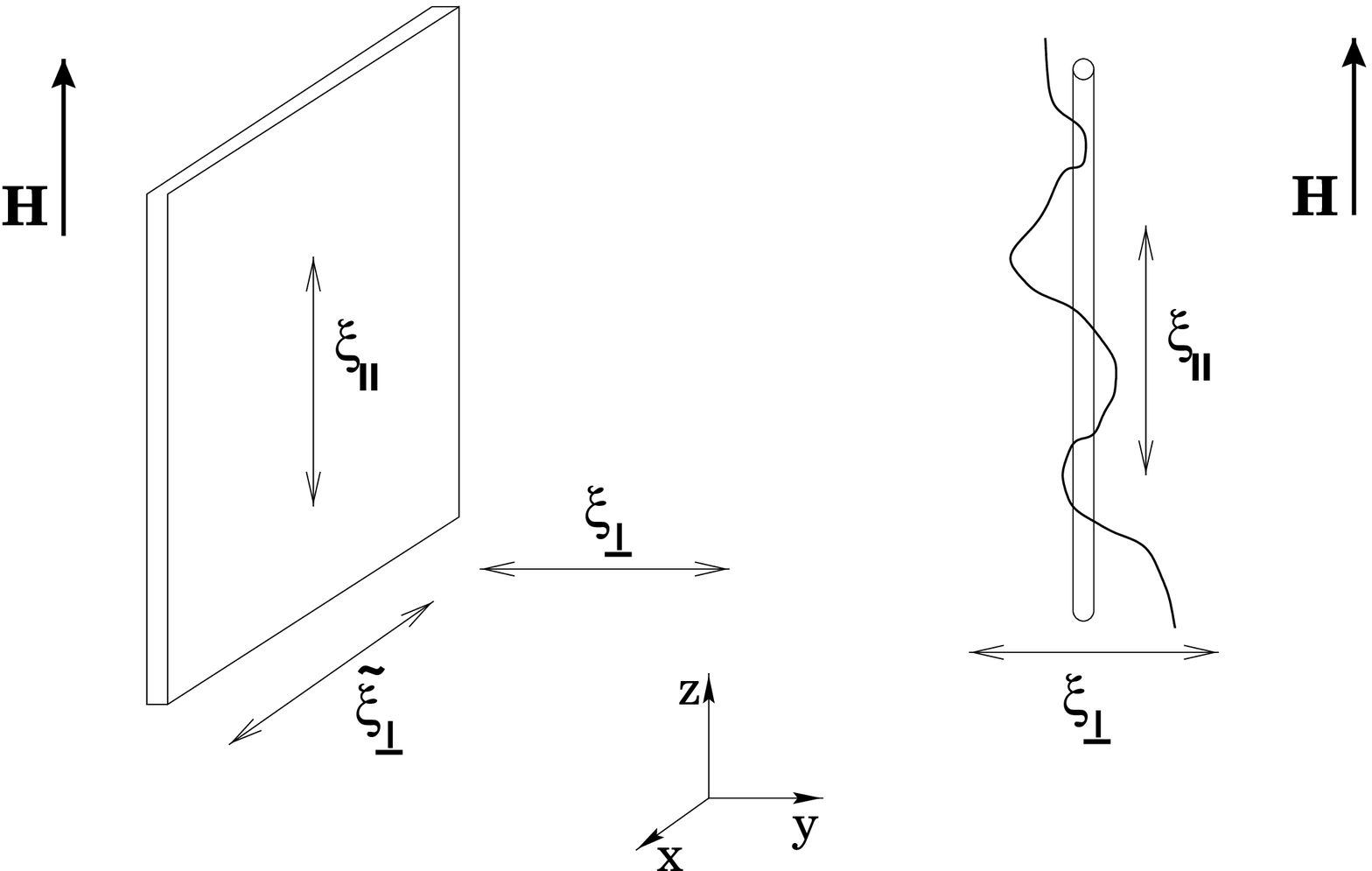}
\label{lengths}
\end{center}
\vspace{0.1in}
\noindent {\bf Figure 1}. The two anisotropic disorder geometries discussed in the text
and the corresponding diverging correlation lengths.
\vspace{0.2in}

Scaling can then be used to relate physical quantities to these diverging correlation
lengths and time. In particular, the linear resistivity $\rho_\perp(T)$ of
the vortex liquid for currents applied in the $ab$ plane
is predicted to vanish as $T\rightarrow T_{G}$ from above as
\begin{equation}
\label{rhoperp} 
\rho_\perp(T)\sim |T-T_{G}|^{\nu(z-\zeta)}
\end{equation}
in bulk samples in three dimensions\cite{ffh,drnvv}.

Although some predictions of the scaling theory  have been tested experimentally,
there are as yet no direct measurements of the transport coefficients usually associated
with glass transitions in conventional forms of matter, such as the shear viscosity.
In this paper we show that the liquid viscosities also exhibit
strong divergences at a continuous disorder-driven glass transition of vortex matter.
We make explicit predictions for the critical exponents controlling the divergence
of the liquid viscosities
and propose experiments which test 
our predictions. 
For instance, the behavior of the shear viscosity is found to be determined by
the dynamical critical exponent $z$ that controls the divergence of the relaxation
time in the  glass phase. A direct measurement of the shear viscosity
would therefore provide a direct probe of the diverging relaxation time
associated with glassy behavior.\cite{kes} A brief account of this work has been
presented elsewhere. \cite{BGshort}

As for ordinary matter, the shear rigidity of the vortex array can be probed by driving 
the vortices to flow in confined geometries. \cite{mcmdrn} The fabrication and use of such confined
geometries was pioneered by Kes and collaborators \cite{pruijm,kes} to study the shear rigidity of the
two-dimensional vortex liquid near the freezing transition in thin
superconducting films. More recently,
patterned irradiation of cuprate superconductors with heavy ions has made it possible
to create samples with controlled distributions of damage tracks that will
allow for a new 
generation of experiments that may in fact provide a direct probe of viscous
critical behavior
near the Bose glass transition \cite{pastoriza}. Specifically, by starting with a clean sample,
at temperatures such that point disorder is negligible,
it is possible to selectively  irradiate  regions
of controlled geometry
by covering the top of the sample with a suitable mask. 
Perhaps the simplest geometry that can be created is that of the channel shown in Fig.~2.
This closely parallels an experimental setup proposed several years ago 
for measuring the shear viscosity
of an entangled vortex liquid forced to flow in 
the channel between two parallel twin planes. \cite{mcmdrn} Now we propose that a similar geometry created 
by selective heavy ion irradiation may be used to probe the growing correlations
at the Bose glass transition.\cite{BGshort}
\begin{center}
\vspace{0.2in}
\epsfxsize=3.5in
\epsfbox{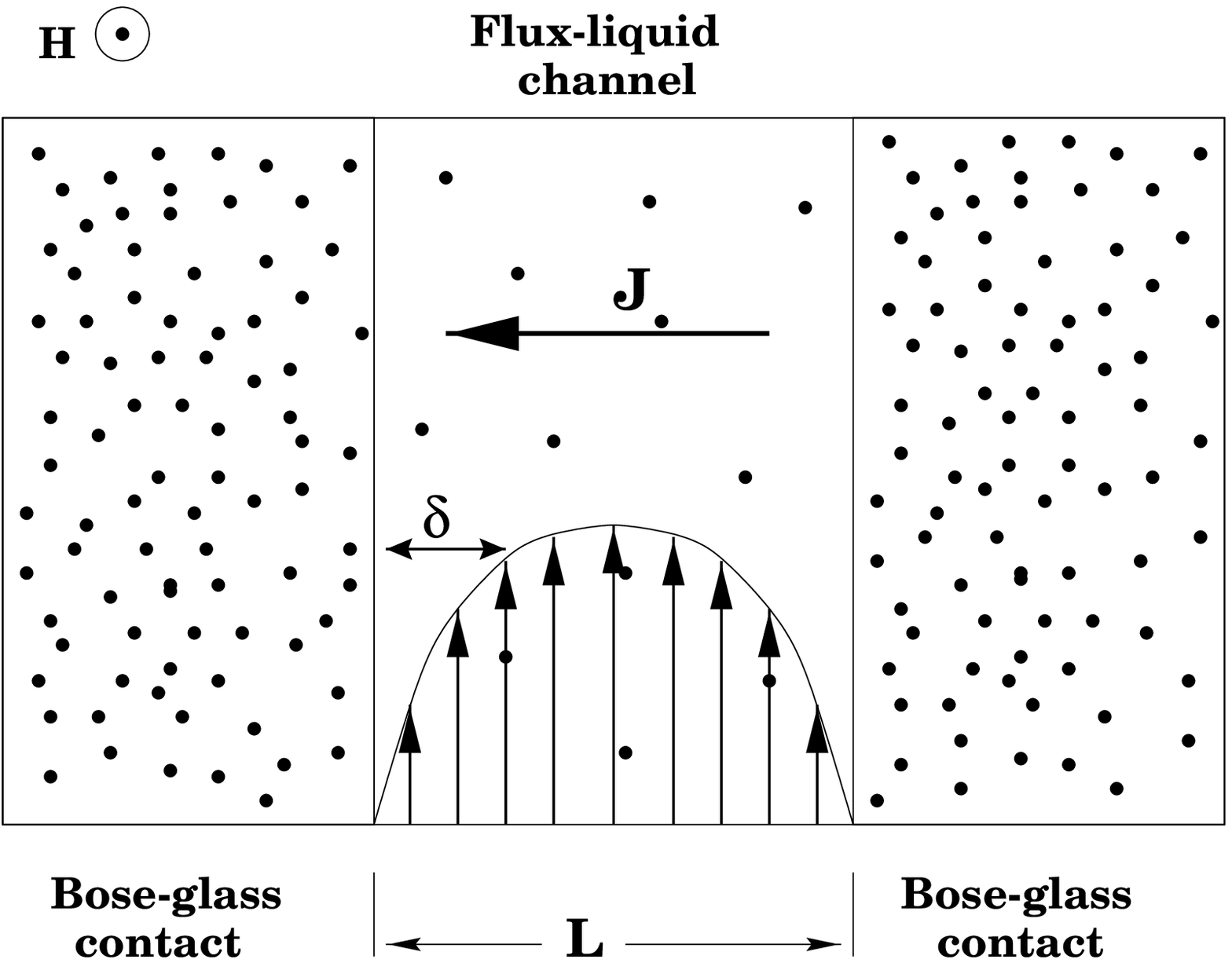}
\label{channel}
\end{center}
\vspace{0.1in}
\noindent {\bf Figure 2}. A weakly irradiated channel where the flux liquid is sandwiched between 
two heavily irradiated Bose-glass contacts. 
A current $J$ applied across the channel yields flux motion
along the channel. Vortex pinning at the boundaries propagates a length $\delta$ into the channel,
yielding a spatially inhomogeneous flow profile.
\vspace{0.2in}

\noindent As shown in Fig.~2,
the side regions of the samples have been heavily irradiated, and are characterized by a high matching field
$B_{\phi}^{(2)}$ and transition curve $T_{BG}^{(2)}$, 
while the channel is lightly irradiated with a lower matching field
$B_{\phi}^{(1)}<B_{\phi}^{(2)}$ and transition curve $T_{BG}^{(1)}$. 
When $T_{BG}^{(1)}<T_{BG}<T_{BG}^{(2)}$, the flux array in the channel
is in the liquid state, while the contacts are in the Bose glass phase.
Flow in the resistive
flux liquid region is impeded by the ``Bose-glass contacts'' at the boundaries, as the many trapped vortices in
these regions provide an essentially impenetrable barrier for the flowing vortices, due to
their mutual interactions.
As the temperature is lowered at constant field, so that
the Bose glass transition $T_{BG}^{(1)}$ of the liquid region is approached from above
(Fig.~3) the Bose glass
correlation length increases, forcing the pinning at the boundaries to propagate 
into the liquid channel and yielding a spatially inhomogeneous electric field profile
which can be probed experimentally. We will show that the spatial inhomogeneity of the electric field 
occurs precisely on length scales given by the correlation length,
providing an unambiguous prescription for extracting critical exponents from 
this type of transport measurements.

By controlling the disorder in the channel region, the same geometry can also be used to
study the onset of rigidity in the vortex array near the vortex glass or the 
polymer glass transitions. In the former case the channel (now free of damage tracks)
should contain a high concentration of point defects (augmented, perhaps, by proton irradiation
\cite{lopez}). The polymer glass transition
may take place when the channel is ``clean'', with only a low concentration of oxygen 
vacancies. A crucial difference between the polymer glass transition and the 
continuous disorder-driven glass transitions discussed above is that the former is a 
nonequilibrium phenomenon associated with the slowing down of the system dynamics on experimental
time scales and does not exhibit universal critical behavior.
For this reason, by testing for a divergent length
scale, transport measurements in confined geometries may actually
be used to {\it distinguish} experimentally between a continuous disorder-driven glass transition and
the polymer glass transition.\cite{nelson_review}
\vspace{0.2in}
\begin{center}
\epsfxsize=3.0truein
\epsfbox{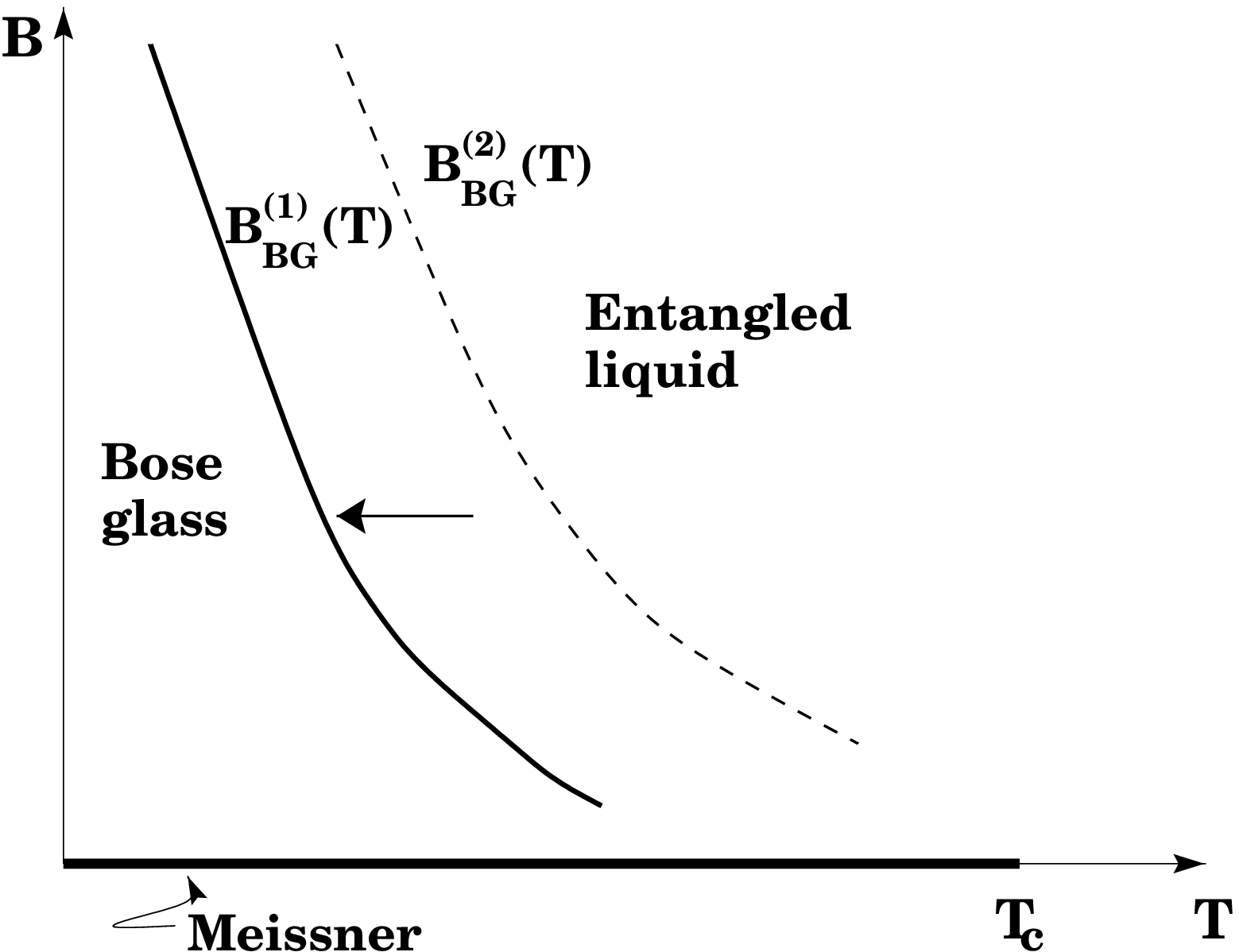}
\label{PhaseDiag}
\end{center}
\vspace{0.2in}

\noindent {\bf Figure 3}. A sketch of the $(B,T)$ phase diagram for the flux array in the weakly irradiated channel region.
The heavy line, $B_{BG}^{(1)}(T)$, denotes the continuous transition from the Bose glass to the entangled liquid.
Also shown is the location $B_{BG}^{(2)}(T)$ of the Bose glass transition line in the heavily irradiated 
contacts. When a field $B_{BG}^{(1)}(T)<B<B_{BG}^{(2)}(T)$ is applied, the flux array in the channel is in the liquid state,
while the contacts are in the Bose glass phase. By decreasing the temperature at constants field along
the direction of the arrow, the Bose glass transition of the channel region is approached from 
above. When $\overline{B}$, instead of $H$, is used for the phase diagram, the Meissner 
phase collapses to a line, as shown.
\vspace{0.2in}
%
%

As shown in Ref. \onlinecite{mcmdrn}, the electrodynamics
of vortex liquids is in general  nonlocal due to interactions and entanglement, even away from phase transitions.
Such nonlocalities may become very important near a phase transition, particularly when
the transition is continuous.
A natural way of incorporating the nonlocality in the long wavelength description of vortex liquid
dynamics is via a set of hydrodynamic equations where the effect of 
intervortex interactions appears as 
viscous forces. A complete set of hydrodynamic equations for the vortex liquid is given in section \ref{sec:hydro}.
These equations incorporate driving forces from external current 
both parallel and transverse to the field ${\bf H}$, as well as both compressional forces and pressure forces 
due to vortex segments that are not aligned with the $z$ axis. They therefore generalize earlier hydrodynamic
equations written down by us that only considered flow driven by currents in the $ab$ plane.
\cite{mcmdrn}
As pointed out earlier by Mou et al. \cite{mou}, the line nature of vortex matter yields to a proliferation of
viscosity coefficients, much like in liquid crystals. There are therefore six independent 
vortex liquid viscosities in these equations. The general equations are complicated and will be discussed 
in more detail below. The equation for the coarse-grained flow velocity ${\bf v}({\bf r})$ of the vortex liquid
 in the channel geometry of Fig.~2 is, however, very simple, namely, \cite{mcmdrn}
\begin{equation}
\label{v_channel}
-\gamma_\perp{\bf v}+\eta\nabla^2_\perp{\bf v}={1\over c}n_0\phi_0\hat{\bf z}\times{\bf J},
\end{equation}
where $\gamma_\perp$ is the friction coefficient that incorporates both the interaction of the vortex cores with 
the underlying ionic lattice and the pinning by material defects and $\eta$ is the vortex liquid shear viscosity.
The term on the right hand side of Eq. (\ref{v_channel}) is the Lorentz current due to an external 
current ${\bf J}$ applied across the channel, with $n_0$ the areal density of vortices. 
Since the local electric field from flux motion is ${\bf e}({\bf r})={n_0\phi_0\over c}\hat{z}\times{\bf v}({\bf r})$,
Eq. (\ref{v_channel}) can be rewritten as a nonlocal version of Ohm's law,
\begin{equation}
\label{E_channel}
{\bf e}-\delta^2\nabla_\perp^2{\bf e}=\rho_\perp{\bf J},
\end{equation}
where 
\begin{equation}
\label{rhoperp2}
\rho_\perp(T)=\Big({n_0\phi_0\over c}\Big)^2{1\over\gamma_\perp(T)}
\end{equation}
is the in-plane linear resistivity in a bulk sample.
The viscous length 
\begin{equation}
\delta(T)=\sqrt{\eta(T)/\gamma_\perp(T)}
\end{equation}
controls the importance of viscous 
drag. In a bulk sample and in the absence of strong inhomogeneities in the spatial
distribution of disorder, the electrodynamic response is expected to be essentially 
local on large scales. Far from the boundaries in macroscopic samples,  
the viscous drag drops out from Eq. 
(\ref{E_channel}) and the linear vortex liquid resistivity is given by Eq. 
(\ref{rhoperp2}) and is controlled entirely by the friction coefficient, 
$\gamma_\perp$. In contrast, in a confined geometry viscous drag becomes
dominant near a glass transition and the viscous length controls the flux liquid
resistivity in this case.

For continuous disorder-driven glass transitions, such as the vortex or Bose glass
transitions, a dimensional analysis of the hydrodynamic equations, combined with the scaling hypothesis that the diverging correlation lengths $\xi_\perp$ and $\xi_\parallel$ are
the only relevant length scales in the problem, leads to the identification of the viscous length with the in-plane correlation length,
\begin{equation}
\label{length}
\delta(T)\sim\xi_\perp(T)\sim|T-T_{G}|^{-\nu}.
\end{equation}
Since the bulk resistivity vanishes at $T_{BG}$ according to Eq. (\ref{rhoperp}), or equivalently the friction 
$\gamma_\perp(T)$ diverges with the same exponent,
\begin{equation}
\label{gammaperp}
\gamma_\perp(T)\sim|T-T_{BG}|^{-\nu(z-\zeta)},
\end{equation}
the identification of these two length scales
immediately leads to the prediction that the liquid shear viscosity diverges at a
continuous glass
transition as 
\begin{equation}
\label{eta}
\eta(T)\sim\xi_\perp^2(T)\gamma_\perp(T)\sim |T-T_{BG}|^{-\nu(z+2-\zeta)}.
\end{equation}
At the Bose glass transition, where $\zeta=2$ and $\nu\simeq 1$, the divergence of the shear viscosity is controlled
by the dynamical exponent $z$.

In contrast, there is no obvious  diverging correlation length controlling the polymer glass transition
proposed some time ago by one of us \cite{nelsonprl}
as an alternative to the vortex glass scenario and 
a possible explanation for the irreversibility line observed experimentally at high
fields. In this scenario a viscous vortex liquid with large barriers to line crossing,
when cooled rapidly may drop out of equilibrium at a polymer glass
transition temperature $T_{PG}$, well before it has time to nucleate  the 
equilibrium crystalline phase. The polymer glass is metastable and the transition is
analogous to the glass transition conjectured in ordinary forms of matter.
Disorder plays no role in the transition itself (although it may have dramatic
effects in controlling the dynamical response of the vortex array) and the vortex
friction $\gamma_\perp$ remains finite across the transition.
The shear viscosity grows rapidly as $T\rightarrow T_{PG}^+$ and its temperature dependence
might conceivably be described by the standard
Vogel-Fulcher form, 
\begin{equation}
\label{VF}
\eta(T)\approx\eta_0e^{c/(T-T_{PG})},
\end{equation}
which works very well
for glass transitions in conventional forms of matter. Experiments 
capable of measuring both the friction coefficient $\gamma_\perp$ (which can be extracted
from a bulk resistive measurements) and the viscous length
$\delta$ (which can be obtained from resistive measurements 
in confined geometries) can therefore distinguish between a kinetic polymer glass transition and
a continuous-disorder driven glass transition, as well as providing direct measurements
of the critical exponents in the latter case.\cite{nelson_review}
At a continuous glass transition, both friction and viscosity will diverge
 with a characteristic critical exponent.
At a polymer glass transition, the friction will remain finite and the viscosity is 
expected to exhibit the Vogel-Fulcher behavior given by Eq. (\ref{VF}).

As shown in section \ref{sec:scaling}, dimensional analysis of the general  hydrodynamic equations
yields predictions for the scaling exponents for all the six vortex liquid viscosities
near a continuous glass transition.
The results (see Sec. II for the precise definition of these quantities) are summarized in table 1.
\vspace{0.2in}
\begin{center}
\begin{tabular}{|c|c|}\hline
$\gamma_\perp$ & $\nu(z-\zeta)$\\\hline
$\gamma_\parallel$ & $\nu (z-2+\zeta)$\\\hline
$\eta$ & $\nu(z+2-\zeta)$\\\hline
$\eta_b$ & $\nu(z+2-\zeta)$\\\hline
$\eta_z$ & $\nu(z+\zeta)$\\\hline
$\eta^t$ & $\nu(z+\zeta)$\\\hline
$\eta^t_z$ & $\nu(z-2+3\zeta)$\\\hline
$\eta_x$ & $\nu(z+\zeta)$\\\hline
\end{tabular}
\end{center}
\vspace{0.2in}
{\bf Table 1.} A summary of the critical exponents for the friction and viscosity
coefficients (see Sec. II for definitions) near a continuous glass transition in three dimensions. 
Each transport coefficient diverges at $T_{BG}$ as $\sim t^{-\alpha}$,
where $t=|T-T_{BG}|/T_{BG}$. The value of $\alpha$ for each transport coefficient is 
indicated in the second column. The anisotropy exponent $\zeta$ should be chosen as 
$\zeta=2$ for the Bose glass transition and $\zeta=1$
for the vortex glass transition.

\vspace{0.3in}

We stress that the  identification of the {\it dynamical} length scale $\delta$ with the 
{\it static} correlation length $\xi_\perp$ is by no means  obvious.
It mirrors the corresponding result obtained in the theory of continuous melting in two dimensions,
where the liquid shear viscosity scales as $\eta\sim 1/\xi_d^2$, with $\xi_d$ the average 
distance between unbound dislocations, which in turn diverges at the transition, yielding a corresponding 
divergence of the viscosity. \cite{drn_domb} In this case the precise relationship between static and dynamical properties
is one of the outcomes of the theory of two-dimensional melting which occurs via successive dislocation  and disclination
unbinding.
For the vortex or Bose glass transitions the divergence of the correlation length 
is due to collective effects mediated by extrinsic quenched disorder. The assumption made 
here is that this growing correlations associated with disorder will also drive the divergence
of dynamical properties in the liquid phase, namely the viscosities. 

This hypothesis has powerful implications. As shown in Ref. \onlinecite{BGshort}
and described in more detail in section \ref{sec:confined} below, by
combining scaling in spatially inhomogeneous geometries with hydrodynamics we are able to
obtain not just the scaling exponents for the transport coefficients, but also 
the full form of the scaling function for the linear resistivity of the vortex liquid in
a constrained geometry. The precise form of the scaling function depends on the details
of the geometry and was given in \onlinecite{BGshort} for the channel and for
the Corbino disk
geometry (Fig.~4) used recently by L\'opez et al. \cite{lopez}.
Once the form of the scaling function is known, a quantitative analysis of transport experiments 
can in principle be carried out to extract the critical exponents.

The plan of the paper is as follows. In section \ref{sec:hydro} we generalize the hydrodynamic equations 
of Ref.  \onlinecite{mcmdrn} to incorporate gradients and driving forces in both the directions parallel
and transverse to the applied field. Upon neglecting the small transverse frictional forces responsible for the Hall
effect, the hydrodynamic equations contain two friction coefficients and six viscosity coefficients.
In section \ref{sec:scaling}, after briefly summarizing the scaling theory for the Bose glass transition in bulk,
we discuss the dimensional analysis of the hydrodynamic equations and obtain the critical exponents for all
the transport coefficients, as given in table 1. In section \ref{sec:confined} we combine scaling in spatially inhomogeneous
systems with hydrodynamics to show how the full form of the scaling function can be obtained for specific
geometries. The Corbino disk geometry is discussed in some detail. Finally, section V 
generalizes both the scaling theory and the hydrodynamics to finite frequency to discuss the 
critical scaling of the 
linear response of the vortex liquid to  ac drives at the Bose glass transition.
Appendix A makes contact with the earlier work
by Mou et al. \cite{mou} on nonlocal effects in vortex liquids. In appendix B we apply scaling ideas to the analysis of transport in the dc flux transformer geometry.

\section{Hydrodynamics of Flux-Line Liquids}
\label{sec:hydro}

Here we generalize the hydrodynamics of flux-line liquids discussed earlier by us \cite{mcmdrn} to
incorporate the effect of driving currents in arbitrary directions, with components both 
parallel and perpendicular to the external field ${\bf H}_0={\bf \hat{z}}H_0$ responsible for the onset of
the vortex state. The resulting hydrodynamic equations contain the six viscosity coefficients 
discussed by Mou et al. \cite {mou} that control the nonlocal electrodynamic response of flux liquids.

The electrodynamics of a type-II superconductor in the mixed state is described by Maxwell's
equations for the local fields ${\bf b}({\bf r},t)$ and ${\bf e}({\bf r},t)$, 
\begin{eqnarray}
\label{maxwell_b}
& &{1\over c}\partial_t{\bf b}+\bbox{\nabla}\times{\bf e}=0,\\
\label{maxwell_e}
& &\bbox{\nabla}\times{\bf b}={4\pi\over c}{\bf j},
\end{eqnarray}
where we have dropped  the displacement current which is negligible at all but 
very high frequencies ($\omega\simeq c/\lambda\sim 10^{15}{\rm Hz}$) 
and ${\bf j}$ denotes the total current density, including the 
equilibrium response of the medium, in addition to any external current. 
In addition, the fields satisfy divergence equations,
\begin{eqnarray}
\label{cont_b}
& & \bbox{\nabla}\cdot{\bf b}=0,\\
\label{cont_e}
& & \bbox{\nabla}\cdot{\bf e}=4\pi\rho.
\end{eqnarray}
Both fields can be obtained from a vector potential ${\bf A}$ and a scalar potential $\phi$, according to 
${\bf b}=\bbox{\nabla}\times{\bf A}$ and
${\bf e}=-{1\over c}\partial_t{\bf A}-\bbox{\nabla}\phi$. We work here with the 
Coulomb gauge,
$\bbox{\nabla}\cdot{\bf A}=0$.
The equations for the fields must be supplemented with a constitutive equation for the current ${\bf j}$.
In the linear (Ohmic) regime this is simply given by a nonlocal generalization of Ohm's
law,
\begin{equation}
\label{ohm}
j_\alpha({\bf r})=\int_{\bf r'}\Sigma_{\alpha\beta}({\bf r},{\bf r'})e_\beta({\bf r'}),
\end{equation}
where Greek indices $\alpha, \beta ,...$ run over the values $x,y,z$ and 
are used to denote Cartesian components in three dimensions
and $\Sigma_{\alpha\beta}$ is a nonlocal conductivity tensor. 
Below we will also use Latin indices $i,j,...$ to denote Cartesian coordinates
in the $xy$ plane, i.e., Latin indices will only assume values $x$ and $y$.
Nonlocality must be incorporated
when describing the mixed state of the cuprates where interactions and entanglement
among the vortices can yield long-ranged  correlations and large scale nonlocal 
electrodynamic response.
It is useful for the following to rewrite Eq. (\ref{ohm}) by separating the current
in an external and internal contributions, ${\bf j}={\bf J}+{\bf j}^{\rm int}$, with
${\bf j}^{\rm int}={c\over 4\pi}\bbox{\nabla}\times({\bf b}-{\bf H})$.
Equation (\ref{ohm}) can then be written as
\begin{equation}
\label{ohm_mod}
J_\alpha({\bf r})
-c \Big(\bbox{\nabla}\times{\delta F\over\delta {\bf b}}\Big)_\alpha
=\int_{\bf r'}\Sigma_{\alpha\beta}({\bf r},{\bf r'})e_\beta({\bf r'}),
\end{equation}
where $-4\pi{\delta F\over\delta {\bf b}}={\bf b}-{\bf H}$ represents the local 
magnetization of the medium, with
$F(T,{\bf b})$ the Helmoltz free energy of the superconductor in an external field, 
${\bf H}$.
The second term on the left hand side of Eq. (\ref{ohm_mod}) incorporates 
the static Meissner response of the material as well as
the current due to local Magnus and pressure forces from the vortices.

In this paper we are interested in fluctuations in the ${\bf b}$ and ${\bf e}$ 
fields due to fluctuations in the vortices degrees of freedom. The relationship between 
the local supercurrent, ${\bf j}_s$, and the vortex degrees of freedom 
is obtained by minimizing the 
Ginzburg-Landau free energy functional for fixed vortex configurations.
Note that in general by using a two-fluid picture there will also be a
``normal current'' contribution to the internal current,
${\bf j}^{\rm int}={\bf j}_s+{\bf j}_n$. We will not incorporate 
this normal part of the response here.
In the London approximation, where the magnitude of the superconducting order 
parameter is assumed constant, and only fluctuations in its phase, $\theta$, are
retained, one obtains the usual London equation,
\begin{equation}
\label{london}
\lambda^2{\bf j}_s=-{c\over 4\pi}\big({\bf A}-\phi_0\bbox{\nabla}\theta\big),
\end{equation}
where $\lambda$ is the penetration length and $\phi_0=hc/2e$ the flux quantum.
For clarity here we discuss only the case of an isotropic superconductor.
The equations are easily generalized to a uniaxial material. 
In addition to the field fluctuations described by Eq. (\ref{london}), there are field 
fluctuations representing thermal deviations from the solution of London equation, which are 
neglected here.

In the chosen gauge ($\bbox{\nabla}\cdot{\bf A}=0$) the longitudinal part of the supercurrent
is simply determined by the phase in the usual way,
\begin{equation}
\bbox{\nabla}\cdot{\bf j}_s={c\phi_0\over 4\pi\lambda^2}\nabla^2\theta.
\end{equation}
Vortices provide a source of vorticity for the supercurrent, as 
the curl of Eq. (\ref{london}) gives
\begin{equation}
\label{curljs}
\bbox{\nabla}\times{\bf j}_s=-{c\over 4\pi\lambda^2}\Big({\bf b}-\phi_0\bbox{\bf T}\Big),
\end{equation}
where ${\bf T}$ is the vortex density vector,
\begin{equation}
\label{vort_dens}
{\bf T}({\bf r},t)={1\over 2\pi}\bbox{\nabla}\times\bbox{\nabla}\theta.
\end{equation}
We note that the curl of $\bbox{\nabla}\theta$ is nonzero in the presence of vortices.
By combining Eq. (\ref{curljs}) with Eq.
(\ref{maxwell_e}), we immediately obtain the London equation 
determining the field due to the vortices,
\begin{equation}
\label{london_b}
{\bf b}-\lambda^2\nabla^2{\bf b}=\phi_0{\bf T},
\end{equation}
To obtain the contribution to the electric field from the vortex degrees of freedom,
we differentiate Eq. (\ref{london}) with respect to time and use 
${\bf e}=-{1\over c}\partial_t{\bf A}-\bbox{\nabla}\phi$, with the result
\begin{equation}
\label{london_e}
{\bf e}=-\bbox{\nabla}\phi+{4\pi\lambda^2\over c^2}\partial_t{\bf j}_s
      -{\phi_0\over c}\partial_t\bbox{\nabla}\theta.
\end{equation}
The vortex part of the electric field is then
given by
\begin{eqnarray}
\label{london_eT}
& & {\bf e}^v-\lambda^2\nabla^2{\bf e}^v=
       {\phi_0\over 2c}\epsilon_{\alpha\beta\gamma}Q_{\beta\gamma},
\end{eqnarray}
where 
\begin{equation}
\label{vort_flux}
{1\over 2}\epsilon_{\alpha\beta\gamma}Q_{\beta\gamma}({\bf r},t)
    =-\bigg(\partial_t\partial_\alpha\theta({\bf r},t)\bigg)^T.
\end{equation}
The antisymmetric tensor $Q_{\beta\gamma}$ as defined as in Eq. (\ref{vort_flux})
describes the vortex current. It is precisely the vortex flux tensor as with
the definition (\ref{vort_flux}) the vortex density
vector ${\bf T}$ satisfies the exact conservation law (conservation of vorticity)
\begin{equation}
\label{Tcontinuity}
\partial_tT_\alpha+\partial_\beta Q_{\alpha\beta}=0.
\end{equation}

The focus of the present paper is on the flux-line liquid regime of vortex matter.
In order to take advantage of the vast phenomenology developed for describing long-wavelength
static and dynamical properties of dense liquids with partcle-like
or polymer-like degrees of freedom, we will focus our attention below on the
vortex degrees of freedom, rather that on the fields ${\bf b}$ and ${\bf e}$.
When the $z$ direction is chosen along the external field ${\bf H}$, vortex configurations 
are conveniently parametrized in terms of a set of $N$ single-valued functions ${\bf r}_n(z)$,
which specify the position of the $n$-th vortex line in the $xy$ plane as it traverses 
the sample along the field direction. The three-dimensional position of each flux line is then 
${\bf R}_n(z)=[{\bf r}_n(z),z]$ and the vortex density vector and flux tensor
are given by,
\begin{eqnarray}
\label{T}
& &{\bf T}({\bf r},t)=\sum_{n=1}^N{\partial{\bf R}_n\over\partial z}\delta({\bf r}-{\bf r}_n(z,t)),\\
\label{Q}
& & Q_{\alpha\beta}({\bf r},t)=\epsilon_{\alpha\beta\gamma}\sum_{n=1}^N\Big(
   \partial_z{\bf R}_{n}\times{\bf v}_n\Big)_\gamma\delta({\bf r}-{\bf r}_n(z,t)),
\end{eqnarray}
with ${\bf r}=({\bf r_\perp},z)$ and ${\bf v}_n(z,t)=\partial_t{\bf r}_n$  the vortex
velocity. 
The three-dimensional vortex density vector is also often written as 
\begin{equation}
{\bf T}=({\bf t},n),
\label{Tvector}
\end{equation}
where $n$ is the local areal density of vortices
and ${\bf t}$ describes the local tilt of vortex lines  away from the $z$ direction.
Neglecting spatial inhomogeneities on scales $\leq\lambda$, Eq. (\ref{london_b}) gives
\begin{eqnarray}
\label{b_vortex}
& & b_z({\bf r},t)=\phi_0n({\bf r},t),\nonumber\\
& & {\bf b}_\perp({\bf r},t)=\phi_0{\bf t}({\bf r},t).
\end{eqnarray}
Similarly, the vortex flux tensor $Q_{\alpha\beta}$ can be written
as
\begin{equation}
\left( \begin{array}{ccc}
Q_{xx} & Q_{xy} & Q_{xz}\\
Q_{yx} & Q_{yy} & Q_{yz}\\
Q_{zx} & Q_{zy} & Q_{zz}
\end{array} \right )
=
\left( \begin{array}{ccc}
0 & n_0V &-n_0v_x\\
-n_0V & 0 & -n_0v_y\\
n_0v_x & n_0v_y & 0
\end{array} \right ),
\label{Qmatrix}
\end{equation}
where ${\bf v}$ represent the flow velocity of vortices moving in the $xy$ plane and 
$n_0V={1\over 2}\epsilon_{ij}Q_{ij}$ is the number of vortex segments aligned with the $x$ direction crossing an area 
normal to the $y$ direction per unit area and per unit time, with
\begin{eqnarray}
\label{jv}
& &n_0{\bf v}({\bf r},t)=\sum_{n=1}^N{\bf v}_n\delta({\bf r}-{\bf r}_n(z,t)),\\
\label{Q0}
& & n_0V({\bf r},t)=\sum_{n=1}^N\hat{\bf z}\cdot\Big(
   \partial_z{\bf r}_{n}\times{\bf v}_n\Big)\delta({\bf r}-{\bf r}_n(z,t)).
\end{eqnarray}
We have explicitly linearized in the fluctuations as we are only interested in 
linearized hydrodynamic equations. Neglecting again nonlocalities on length scales $\sim\lambda$,
Eq. (\ref{london_eT}) yields
\begin{equation}
e^v_\alpha={\phi_0\over 2c}\epsilon_{\alpha\beta\gamma}Q_{\beta\gamma},
\end{equation}
or
\begin{eqnarray}
& & e^v_z={\phi_0\over 2c} \epsilon_{ij}Q_{ij}={n_0\phi_0\over c}V,\nonumber\\
& & e^v_i=-{\phi_0\over c}\epsilon_{ij}Q_{zj}=-{n_0\phi_0\over c}\epsilon_{ij}v_j.
\end{eqnarray}

We are interested here in the long-wavelength properties of flux-line liquids on scales large 
compared to $\lambda$ {\it and} to the average intervortex spacing, $a_0=\sqrt{\phi_0/B_0}$, with $B_0$ the equilibrium 
mean value of the field, $\langle{\bf b}\rangle = \hat{\bf z} B_0$. For this purpose we
can abandon the description in terms of the microscopic vortex degrees of freedom ${\bf R}_n(z)$
in terms of coarse-grained or hydrodynamic fields describing fluctuations
in the conserved variables of the system and corresponding to
those collective degrees of freedom whose relaxation rate vanishes in the long wavelength limit.
For the vortex liquid the relevant conserved variables are the three components of
the vortex density, ${\bf T}({\bf r},t)=({\bf t},n)$. The coarse-grained vortex density is defined 
as in Eq.
(\ref{T}), with the $\delta$-function replaced by a smeared-out two-dimensional $\delta$-function,
$\delta_{BZ}({\bf r_\perp})$, with a finite spatial extent of the order of the inverse of the 
Brillouin zone boundary, $k_{BZ}=\sqrt{4\pi n_0}$,
\begin{equation}
\delta_{BZ}({\bf r_\perp})={1\over A}\sum_{q_\perp\leq k_{BZ}}e^{-i{\bf q_\perp}\cdot{\bf r_\perp}}.
\end{equation}

As discussed elsewhere, the long wavelength equilibrium properties of vortex liquids can be described in 
terms of a coarse-grained free energy that fully incorporates the nonlocality of the intervortex interaction.
Such a free energy can be derived by explicit coarse-graining of the microscopic 
vortex energy \cite{mcm_ent} or can be written down phenomenologically using familiar ideas
from liquid state physics. It takes the form of an expansion in the fluctuations of the 
hydrodynamic fields from their equilibrium
values and of the corresponding gradients. Retaining only terms quadratic in the fluctuations,
it is given by
\begin{eqnarray}
F_L={1\over 2 n_0^2}\int_{\bf r}\int_{\bf r'} \Big[& &c_L({\bf r}-{\bf r'})\delta n({\bf r})\delta n({\bf r'})
+c_{44}({\bf r}-{\bf r'}){\bf t}({\bf r})\cdot {\bf t}({\bf r'})\Big],
\label{Fhydro}
\end{eqnarray}
where $\delta n=n-n_0$ and $c_L$ and $c_{44}$ are the nonlocal compressional and tilt elastic moduli
of the flux liquid, given elsewhere.\cite{mcm_ent}

The vortex density vector, ${\bf T}=({\bf t},n)$, satisfies 
an exact conservation law, Eq. (\ref{Tcontinuity}),
or, in terms of $n$ and ${\bf t}$,
\begin{eqnarray}
\label{dens_cont}
& & \partial_t n+\bbox{\nabla}\cdot n_0{\bf v}=0,\\
\label{tilt_cont}
& & \partial_t t_i+\partial_jQ_{ij}=\partial_zn_0v_i,
\end{eqnarray}
with the condition $\bbox{\nabla}\cdot{\bf T}=0$, or
\begin{equation}
\partial_zn+\bbox{\nabla}_\perp\cdot{\bf t}=0,
\label{contz}
\end{equation}
ensuring that flux lines do not start nor stop inside the sample.

To obtain a closed set of hydrodynamic equations it is necessary to specify the 
constitutive equations that express the flux tensor $Q_{\alpha\beta}$ 
in terms of the density field and
its gradients. Since the vortex dynamics is overdamped, the constitutive equation
is simply obtained requiring that the net force on each local volume of flux 
liquid vanishes. Its form is analogous to that of Eq. (\ref{ohm_mod}) and it is given by
\begin{eqnarray}
\label{Qconstitutive}
\tilde{\gamma}_{\mu\nu,\lambda\rho}Q_{\lambda\rho} = \tilde{\eta}_{\mu\nu,\alpha\beta,\lambda\rho}\partial_\alpha
\partial_\beta Q_{\lambda\rho}+\partial_\mu{\delta F_L\over\delta T_\nu}
-\partial_\nu{\delta F_L\over\delta T_\mu}
+{\phi_0\over c}\epsilon_{\mu\nu\lambda}J_\lambda.
\end{eqnarray}
The first term on the left hand side of Eq. (\ref{Qconstitutive}) represents the frictional
force per unit length on the vortex liquid, with $\tilde{\gamma}_{\mu\nu,\lambda\rho}$ a friction tensor.
The second term describes viscous forces from intervortex interaction and entanglement,
with $\tilde{\eta}_{\mu\nu,\alpha\beta,\lambda\rho}$ the viscosity tensor. The third term contains 
Magnus and pressure forces arising from vortex density gradients and the last one is 
simply the Lorentz force due to an external current, {\bf J}.

Neglecting for simplicity transverse drag forces associated with the Hall effect,
there are two independent components of the friction tensor and six viscosity 
coefficients. It is useful to rewrite the constitutive equations in terms of the in-plane
vortex flow velocity ${\bf v}$ and the tilt current $V$, 
\begin{eqnarray}
\label{flux_ab}
\gamma_\perp{\bf v}=\big[\eta\nabla_\perp^2+\eta_z\partial_z^2\big]{\bf v} & &
+\eta_b\bbox{\nabla}_\perp(\bbox{\nabla}_\perp\cdot{\bf v})
-\eta_x\big(\hat{\bf z}\partial_z\times\bbox{\nabla}_\perp\big)V\nonumber\\
& & -n_0\bigg(\bbox{\nabla}_\perp{\delta F_L\over\delta n}-\partial_z{\delta F_L\over\delta{\bf t}}
\bigg) -{n_0\phi_0\over c}\hat{\bf z}\times{\bf J}.
\end{eqnarray}
\begin{eqnarray}
\label{flux_c}
\gamma_\parallel V=\big[\eta^t\nabla_\perp^2+\eta_z^t\partial_z^2\big]V
-\eta_x\hat{\bf z}\partial_z\cdot\big(\bbox{\nabla}_\perp\times{\bf v}\big)
+n_0\hat{\bf z}\cdot\bigg(\bbox{\nabla}_\perp\times{\delta F_L\over\delta{\bf t}}\bigg)
+{n_0\phi_0\over c}J_z.
\end{eqnarray}
The constitutive equations (\ref{flux_ab}) and (\ref{flux_c}) are now written in terms of
forces per unit volume. The coefficients $\gamma_\perp$ and $\gamma_\parallel$ have dimension
of friction per unit volume, as usual, and $\eta$, $\eta_z$, $\eta_b$, $\eta^t$, $\eta^t_z$
and $\eta_x$ have dimensions of fluid viscosities.

The six viscosity coefficients appearing in Eqs. (\ref{flux_ab})  and (\ref{flux_c})
are related to the six components of the nonlocal conductivity discussed by Mou et al.
\cite{mou}
The precise relationship is displayed in Appendix A. Equations (\ref{dens_cont}-\ref{contz})
and (\ref{flux_ab},\ref{flux_c}) provide a closed set of
hydrodynamic equations to describe the response of vortex liquids to external currents and
fields. These equations generalize equations written down earlier by us \cite{mcmdrn}
and by others \cite{huse,mou}.
by incorporating all pressure forces, including those from vortex segments that are
not parallel to the external field direction, all nonlocal viscous forces, as
well as external forces due to driving currents aligned with the external field.
In the next section we will show how  predictions for the singular behavior of for the six viscosities near
a continuous phase transition from the vortex liquid to a glassy state 
can be obtained by simple scaling arguments.

\section{Scaling theory}
\label{sec:scaling}

The divergence of pinning 
energy barriers for vanishing driving currents underlies both the  theories of the Bose 
glass and the vortex glass transitions,
which are expected to be continuous. The properties near the transition have therefore 
been described by a scaling theory in terms of divergent lengths and time scales.
A scaling theory for the transition from a vortex liquid to an isotropic vortex glass was developed by 
Fisher, Fisher and Huse \cite{ffh} and then adapted by Nelson and Vinokur to the transition to
the anisotropic Bose glass. \cite{drnvv}
Here we summarize  this scaling theory in a unified way that applies to both transitions
by keeping track of the anisotropy exponent $\zeta$ that has value $\zeta=2$ for the Bose glass and 
$\zeta=1$ for the vortex glass. Of course the values of the other critical exponents are also 
different in the two cases.

The (Gibbs) free energy density $g$ of the vortex array is a function of temperature $T$ 
and external field, ${\bf H}=\hat{\bf z}H_\parallel +{\bf H}_\perp$, 
with $g=g(T,H_\perp,H_\parallel)$. Here 
$H_\parallel=H_0+\delta H_\parallel$, where $H_0$ is the field responsible for 
setting up the vortex state and the flux lines are on average aligned with the $z$
direction. As predicted by the renormalization group, the singular part of the free
energy density, $g_s$, is assumed to obey a homogeneity relation at the transition
where temperature and the external fields are rescaled by different factors,
\begin{equation}
\label{free_sing}
g_s(t,H_\perp,\delta H_\parallel)=l^{-(d-1+\zeta)}g_s\big(l^{1/\nu}t,l^{\lambda_\perp} H_\perp,
   l^{\lambda_\parallel} \delta H_\parallel\big),
\end{equation}
for arbitrary length rescaling parameter, $l$. Here $t$ is the reduced temperature $t=|T-T_G|/T_G$, with $T_G$
either the Bose glass or the vortex glass transition temperature, and $d$ 
denotes the system dimensionality. The case of interest here is $d=3$, 
corresponding to a bulk superconductor. In this section $\lambda_\perp$ and 
$\lambda_\parallel$ denote critical exponents, not to be confused with penetration lengths. Making the usual choice $l=t^{-\nu}\sim\xi_\perp$, dimensional analysis will allows us 
to relate physical quantities to 
the diverging correlation lengths $\xi_\perp$ and $\xi_\parallel\sim \xi_\perp^\zeta$, as described in Ref. \onlinecite{ffh,drnvv,drnleo,fwgf,lidmar,nverratum}.

The local field in the superconductor is given by
\begin{equation} 
\label{B_def}
{\bf b}=-4\pi {\delta g\over\delta{\bf H}}
\end{equation}
and therefore scales according to
\begin{eqnarray}
\label{B_scaling}
& & b_\perp\sim l^{-(d-1+\zeta+\lambda_\perp)},\\
& & b_\parallel\sim l^{-(d-1+\zeta+\lambda_\parallel)}.
\end{eqnarray}
The longitudinal and transverse components of the response function are
\begin{eqnarray}
& & \delta_{ij}\big(1+4\pi\chi_\perp\big)
   ={\delta^2 g_s\over\delta H_{\perp i}\delta H_{\perp j}}
   =\delta_{ij}{n_0^2\phi_0^2\over c_{44}},\\
& & 1+4\pi\chi_\parallel
   ={\delta^2 g_s\over\delta H_\parallel^2}
   ={n_0^2\phi_0^2\over c_{11}},
\end{eqnarray}
where $\chi_\perp$ and $\chi_\parallel$ are the components of the magnetic 
susceptibility and $c_{44}$ and $c_{11}$ the tilt and compressional elastic moduli
of the vortex array. 
The scaling of the susceptibilities is then given by
\begin{eqnarray}
& & \chi_\perp\sim{1\over c_{44}}\sim l^{-(d-1+\zeta+2\lambda_\perp)},\\
& & \chi_\parallel\sim{1\over c_{11}}\sim l^{-(d-1+\zeta+2\lambda_\parallel)},
\end{eqnarray}

At the vortex glass transition the response of the vortex array
is rotationally invariant and both susceptibilities are expected to remain finite.
This requires
\begin{equation}
\lambda_\perp^{VG}=\lambda_\parallel^{VG}=-(d-1+\zeta)/2.
\end{equation}
The scaling of the local fields at the vortex glass transition is then 
\begin{equation}
\label{B_scaling_VG}
b_\perp\sim b_\parallel\sim {1\over\xi_\perp^{(d-1)/2}\xi_\parallel^{1/2}}
\sim t^{3\nu/2},
\end{equation}
where the last equality applies for $d=3$.

At the Bose glass transition, in contrast, the tilt modulus 
is expected to diverge as $c_{44}\sim l^\zeta$, while $c_{11}$ remains finite.
In this case the response of the system is anisotropic,
with 
\begin{eqnarray}
& &\lambda_\perp^{BG}=-(d-1)/2,\\
& &\lambda_\parallel^{BG}=-(d-1+\zeta)/2,
\end{eqnarray}
and the fields scale as 
\begin{eqnarray}
\label{B_scaling_BG}
& & b_\perp\sim {1\over \xi_\perp^{(d-1)/2}\xi_\parallel}
  \sim t^{3\nu},\nonumber\\
& & b_\parallel\sim {1\over \xi_\perp^{(d-1)/2}\xi_\parallel^{1/2}}
  \sim t^{2\nu},
\end{eqnarray}
where again the last equality in each of Eqs. (\ref{B_scaling_BG}) applies
for $d=3$.

To determine the scaling of fields and currents that control the resistive
properties of the vortex array, we note that
gauge invariance of the  Ginzburg-Landau 
theory requires that the fluctuating vector potential scales according to
\begin{eqnarray}
\label{Ascaling}
& &A_\perp\sim \xi_\perp^{-1},\nonumber \\
& &A_z\sim\xi_\parallel^{-1}.
\end{eqnarray}
The scaling of currents and fields is immediately obtained from their definitions,
${\bf J}=\partial f/\partial{\bf A}$ and ${\bf E}=-(1/c)\partial{\bf A}/\partial t$,
where we denote by ${\rm E}$ the macroscopic (spatially-averaged) field from vortex motion.
It is given by,
\begin{eqnarray}
\label{Jscaling}
& & J_\perp\sim\xi_\perp^{2-d}\xi_\parallel^{-1},\nonumber\\
& & J_\parallel\sim\xi_\perp^{1-d},
\end{eqnarray}
and 
\begin{eqnarray}
\label{Escaling_perp}
& &E_\perp\sim\xi_\perp^{-(1+z)},\nonumber\\
\label{Escaling_par}
& &E_\parallel\sim\xi_\parallel^{-1}\xi_\perp^{-z}.
\end{eqnarray}
A scaling ansatz for the IV characteristic can be written down based on the 
structure of the RG flows as \cite{ffh}
\begin{eqnarray}
\label{Eperp_bs}
& & E_\perp(T,J_\perp)=l^{-(1+z)}E_\perp\Big(l^{1/\nu}t,{l^{(d-2+\zeta)} J_\perp\over ck_BT}\Big),\\
\label{Epar_bs}
& & E_\parallel(T,J_\parallel)=l^{-(\zeta+z)}E_\parallel\Big(l^{1/\nu}t,{l^{d-1} J_\parallel\over ck_BT}\Big).
\end{eqnarray}
With the choice $l=t^{-\nu}\sim\xi_\perp(T)$, we obtain
\begin{eqnarray}
\label{Eperp_fin}
& &E_\perp(1,J_\perp)=\xi_\perp^{-(1+z)}{\cal E}_\perp^\pm
  \Big({\xi_\perp^{d-2}\xi_\parallel J_\perp\over ck_BT}\Big),\\
\label{Epar_fin}
& &E_\parallel(1,J_\parallel)=\xi_\parallel^{-1}\xi_\perp^{-z}{\cal E}_\parallel^\pm
  \Big({\xi_\perp^{d-1} J_\parallel\over ck_BT}\Big),
\end{eqnarray}
where ${\cal E}_{\perp,\parallel}^\pm$ are scaling functions. 
From these one can obtain various results 
described in the literature. In particular, by linearizing Eqs. (\ref{Eperp_bs}) and
(\ref{Epar_bs}) in the ohmic regime, one finds
that both components of the linear flux liquid resistivity vanish as the transition is approached
from the liquid side, according to,
\begin{eqnarray}
\label{rhoperp_bs}
& & \rho_\perp(T)=\Big({n_0\phi_0\over c}\Big)^2{1\over\gamma_\perp(T)}\sim\xi_\parallel\xi_\perp^{-(z+3-d)}
\sim|T-T_{BG}|^{\nu(3+z-d-\zeta)},\\
\label{rhopar_bs}
& & \rho_\parallel(T)=\Big({n_0\phi_0\over c}\Big)^2{1\over\gamma_\parallel(T)}\sim
\xi_\parallel^{-1}\xi_\perp^{-(z+1-d)}\sim|T-T_{BG}|^{\nu(1+z-d+\zeta)}.
\end{eqnarray}
Equivalently, Eqs. (\ref{rhoperp_bs}) and (\ref{rhopar_bs}) give the scaling exponents
controlling the divergence of the longitudinal and transverse friction coefficients, 
$\gamma_\parallel(T)$ and $\gamma_\perp(T)$.

Dimensional analysis of the hydrodynamic equations now allows us to extract predictions 
for the scaling  of the six viscosity coefficients at the transition.
The scaling dimension of the fluctuating density and tilt fields
$n$ and ${\bf t}$ is the same as that of the components of the local field ${\bf b}$ and therefore  
is obtained immediately from Eqs. (\ref{B_scaling_VG}) and (\ref{B_scaling_BG}) for the vortex and Bose glass, respectively.
We scale all length and time scales in the hydrodynamic equations with the correlation lengths
$\xi_\perp$ and $\xi_\parallel$, using the scaling dimension just discussed for the various 
physical quantities appearing in these equations. By requiring the equations to be scale invariant
near the transition, we can then read off the critical scaling of the viscosity coefficients,
\begin{eqnarray}
\label{visc_scaling}
& & \eta\sim\xi_\perp^2\gamma_\perp\sim\xi_\perp^{z+5-d}/\xi_\parallel,\nonumber\\
& & \eta_b\sim\xi_\perp^2\gamma_\perp\sim\xi_\perp^{z+5-d}/\xi_\parallel,\nonumber\\
& & \eta_z\sim\xi_\parallel^2\gamma_\perp\sim\xi_\perp^{z+3-d}\xi_\parallel,\nonumber\\
& & \eta^t\sim\xi_\perp^2\gamma_\parallel\sim\xi_\perp^{z+3-d}\xi_\parallel,\nonumber\\
& & \eta^t_z\sim\xi_\parallel^2\gamma_\parallel\sim\xi_\perp^{z+1-d}\xi_\parallel^3,\nonumber\\
& & \eta_x\sim\xi_\perp^2\gamma_\parallel\sim\xi_\parallel^2\gamma_\perp\sim
\xi_\perp^{z+3-d}\xi_\parallel.
\end{eqnarray}
The critical exponents for all the friction and viscosity coefficients are summarized in Table 1.

\section{Scaling and hydrodynamics in confined geometries}
\label{sec:confined}

The flow of vortex liquids in confined geometries provides a powerful experimental
tool to probe the shear resistance of vortex matter. Experiments of this type have been 
carried out by the group of Peter Kes in thin superconducting films.\cite{kes,pruijm} 
In this two-dimensional case the vortex array 
is expected to melt from a solid into a liquid via the dislocation unbinding mechanism.\cite{drn_domb}
The shear modulus of the lattice vanishes continuously at the transition and
is replaced by a shear viscosity in the liquid phase. The shear viscosity diverges as the 
transition is approached from the liquid side. In order to probe this behavior, Theunissen et al.
studied the flow of a two-dimensional vortex solid through narrow channels in a double-layer device,
consisting of a weakly pinning 
amorphous $Nb_3Ge$ bottom layer, covered with a strongly pinning $NbN$ thick layer. 
Using nanolithography, parallel channels were etched through the $NbN$ layer into the
$NbGe$ layer. By applying a current normal to the direction of the channels, vortices where
then forced to flow along the weakly pinning channels, while the $NbN$ regions
provide the  strong-pinning boundaries. \cite{kes}
By fitting their data to the form for the flux-flow resistivity of a vortex liquid in a channel 
obtained earlier by us via a simple hydrodynamic model\cite{mcmdrn}, these authors were able to extract the 
temperature dependence of the shear viscosity near melting, finding a behavior in agreement with 
the predictions of the theory of two-dimensional melting. \cite{drn_domb}

As discussed in the Introduction, patterned irradiation of cuprate superconductors with heavy ions
allows for experiments similar in spirit to the one carried out by Theunissen et al. \cite{kes}.
These experiments can probe the growing shear viscosity of the vortex array near a 
continuous glass transition and may shed some light on the connection 
between static and dynamical properties of
these glassy systems. 

Scaling theory can again be used to determine the temperature dependence of fields and currents in this spatially
inhomogeneous situation arising from the confined geometry.
A generalized homogeneity relation for the local electric field 
$e_\perp$ perpendicular to any correlated disorder from flux motion at position $x$ in a channel of
thickness $L$ takes the form
\begin{equation}
\label{scaling}
e_\perp(T,J_\perp,x,L)=l^{-(1+z)}e_\perp \bigg(l^{1/\nu}t,{l^{\nu(1+\zeta)}J_\perp\phi_0\over ck_BT},
{x\over l},{L\over l}\bigg),
\end{equation}
By choosing again $l=t^{-\nu_\perp}\sim \xi_\perp(T)$ we obtain
\begin{equation}
\label{scaling2}
e_\perp(T,J_\perp,x,L)=\xi_\perp^{-(1+z)}{\cal E}_\perp\Big({\xi_\perp \xi_\parallel J_\perp\phi_0\over ck_BT},{x\over \xi_\perp},{L\over \xi_\perp}\Big).
\end{equation}
In the entangled flux liquid the response is linear at small current.
Upon expanding the right hand side of Eq. (\ref{scaling2}) we obtain for $J_\perp\rightarrow 0$
\begin{equation}
\label{linear_scaling}
e_\perp(J_\perp\rightarrow 0,x,L)\simeq \rho_\perp^0 \Big({\xi_\perp\over a_0}\Big)^{2-z} J_\perp {\cal F}(x/\xi_\perp,L/\xi_\perp),
\end{equation}
where $\rho_\perp^0=\Big(n_0\phi_0/c\Big)^2(1/\gamma^0_\perp)$ is the Bardeen-Stephen resistivity of noninteracting
flux lines, with $\gamma_\perp^0$ a bare friction. 
A scaling form for the resistivity $\rho_\perp(T,L)=\Delta V/(LJ_\perp)$, with $\Delta V$ the net voltage drop across the channel,
is easily obtained by integrating Eq. (\ref{linear_scaling}), with the result.
\begin{equation}
\label{rho}
\rho_\perp(T,L)=\rho_\perp(T)f(L/\xi_\perp)
\end{equation}
with $f(x)={1\over x}\int_{x/2}^{x/2} du{\cal F}(u,x)$ a scaling function and $\rho_\perp(T)$
the bulk flux liquid resistivity given in Eq. (\ref{rhoperp}).
Here $\gamma_\perp$ is the renormalized  friction coefficient that
incorporates the growing Bose glass correlations near $T_{BG}$,
$\gamma_\perp=\gamma_\perp^0\Big({\xi_\perp\over a_0}\Big)^{z-2}$,
and diverges at the transition as 
$\gamma_\perp\sim|T-T_{BG}|^{-\nu(z-2)}$ \cite{drnvv}.
For $L\gg \xi_\perp$, the channel geometry has no effect and one must recover the bulk result.
This requires $f(x\gg 1)\sim 1$.

The scaling function ${\cal F}$ can be determined by  assuming that
the long wavelength electric field of Eq. (\ref{linear_scaling})
is described
by the set of hydrodynamic equations discussed in Ref. \onlinecite{mcmdrn}.
For simple geometries where the current is applied in the $ab$ plane and the flow is spatially
homogeneous in the $z$ direction, these reduce to the single equation 
for the coarse-grained flux liquid flow velocity ${\bf v}({\bf r})$ given in Eq.
(\ref{v_channel}).
Intervortex interaction at the Bose-glass boundaries yields ``adhesion''
of the flux liquid to these boundaries, forcing the flux flow velocity to vanish or become very small. 
In the hydrodynamic model this translates into a no-slip boundary
condition for the flux liquid velocity at the boundaries. By preventing the free flow of flux liquid,
the Bose glass boundaries can significantly decrease the macroscopic flux-flow resistivity of
the superconductor.
The electric field profile is obtained by solving Eq. (\ref{E_channel}) for the appropriate geometry
and with suitable boundary conditions.

\subsection{Channel geometry}

The solution of the hydrodynamic equation 
for the simple channel geometry sketched in Fig.~2, with a homogeneous current
${\bf J}=-\hat{\bf x} J_\perp$ applied across the channel has been given elsewhere 
\cite{mcmdrn,BGshort} and is displayed here for completeness.
It is given by
\begin{equation}
\label{Echannel}
e_\perp(x,L)=\rho_\perp J_\perp \bigg[1-{\cosh(x/\xi_\perp)\over \cosh(L/2\xi_\perp)}\bigg].
\end{equation}
Upon  comparing Eq. (\ref{Echannel}) to Eq. (\ref{linear_scaling}), we identify the quantity in square
brackets in Eq. (\ref{Echannel}) with the scaling function ${\cal F}$.
The scaling form for the resistivity is obtained by
integrating Eq. (\ref{Echannel}), with the result
\begin{equation}
\rho_{\perp L}(T,L)=\rho_\perp(T)\Big[1-{2\xi_\perp\over L}\tanh\Big({L\over 2\xi_\perp}\Big)
\Big].
\end{equation}
If $\xi_\perp\ll L$, we recover the bulk result given by Eq. (\ref{rhoperp_bs}),

\begin{equation}
\rho_{\perp L}(T,L)\simeq\rho_\perp(T)\sim t^{\nu(z-\zeta)}.
\end{equation}
Near the transition, however, $\xi_\perp\gg L$, and the resistivity depends on the channel width and is
controlled by the shear viscosity, with
\begin{equation}
\label{rhopL}
\rho_{\perp L}(T,L)\simeq {\rho_\perp L^2\over 12 \xi_\perp^2}=
\Big({n_0\phi_0\over c}\Big)^2{L^2\over 12\eta(T)}\sim L^2 t^{\nu (z+2-\zeta)}.
\end{equation}
The strong divergence of the viscosity implicit in Eq. (\ref{rhopL})is precisely the kind of behavior
expected at a liquid-glass transition. The Bose glass transition 
is an example of a glass transition that is well understood theoretically and 
where precise predictions are available.

\subsection{Corbino disk}
Experiments in patterned geometries near the Bose glass transformation provide
an exciting opportunity to probe viscous behavior near a second order glass transition.
Other patterned geometries can be designed and 
used to probe some or all of the other viscosity coefficients. Of particular interest
is the Corbino disk geometry, recently used by L\'opez et al. for defect-free 
materials \cite{lopez} to reduce boundary effects in the flux flow measurements.
This geometry was briefly discussed in \onlinecite{BGshort} and is sketched in Fig.~4.

\vspace*{0.2in}
\epsfxsize=3.0in
\epsfbox{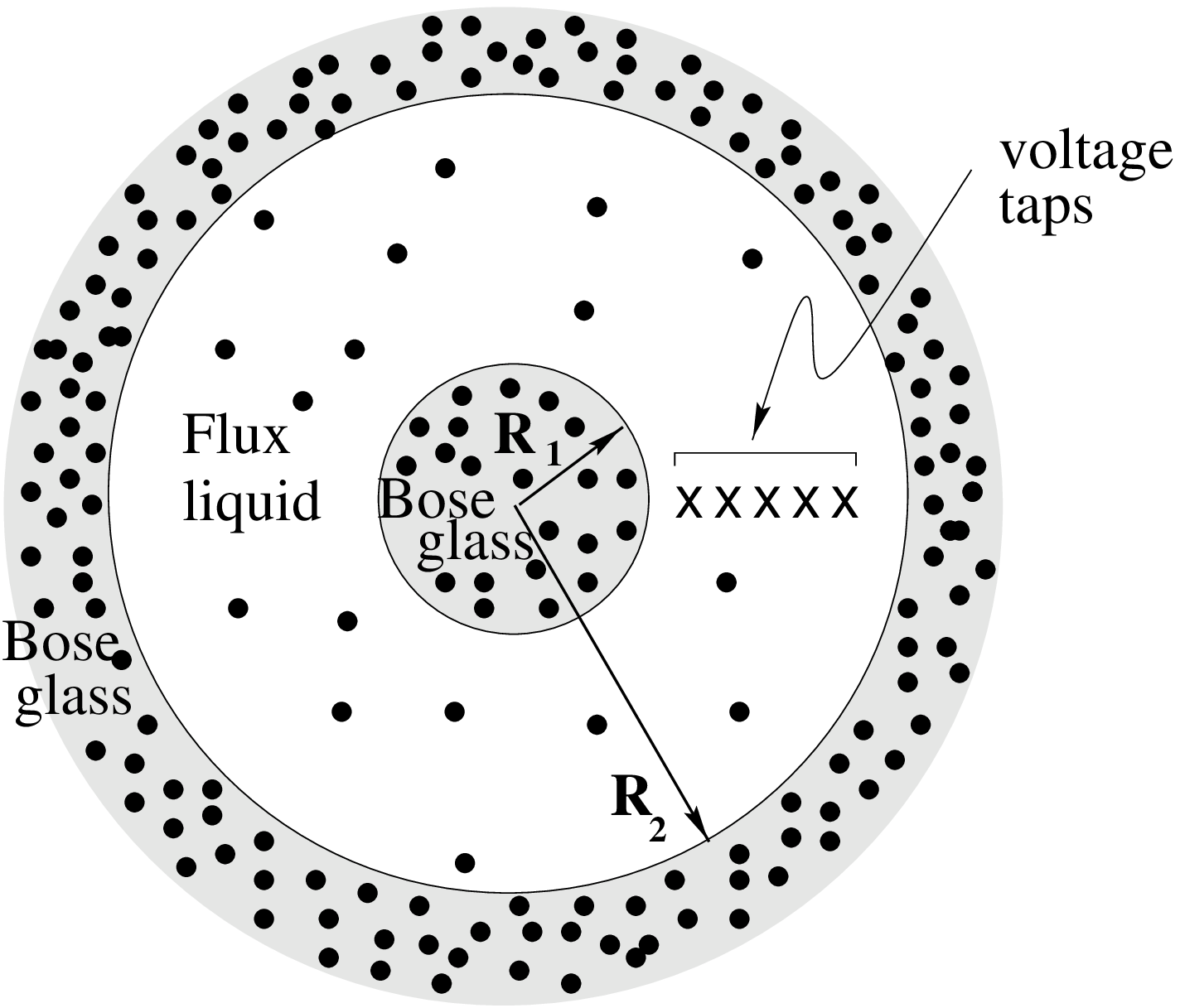}
\label{corbino_topview}
\vspace{0.2in}

\noindent {\bf Figure 4}. Top view of the Corbino disk geometry with Bose glass contacts. The magnetic field is
out of the page. The vortex array is in 
the Bose glass state in the inner and outer densely dotted regions and in the flux liquid state 
in the weakly irradiated annular region. A radial driving current drives flux motion in the azimuthal direction.
\vspace{0.3in}

In \onlinecite{BGshort} we proposed the fabrication of a 
Corbino disk with Bose glass inner and outer contacts. The annular region 
corresponding to $R_1\leq r\leq R_2$ is in the vortex liquid state.
A constant current $I$ is injected at the inner boundary of the cylindrical flux liquid region and extracted at the
outer boundary. The resulting current density is radial, with
\begin{equation}
\label{radj}
{\bf J}={I\over 2\pi W}{\hat{\bf r}\over r},
\end{equation}
with $W=R_2-R_1$ the width of the annular flux liquid channel.
This current induces tangential flow of the vortices,
which in turn yields a radial voltage drop that can be probed by a suitable set of voltage taps, placed
as sketched in Fig.~4. 
If the intervortex spacing $a_0$ in the flux liquid
is small compared to the size $R_2-R_1$ of the annular channel  where flux flow occurs, the flow can be described
by the hydrodynamic equation (\ref{E_channel}). Using cylindrical coordinates with the $z$ axis directed
along the direction of the external field ${\bf H}\parallel c$,
the vortex flow velocity is in the azimuthal direction, 
${\bf v}=v_\phi(r)\bbox{\hat{\phi}}$, corresponding to a radial field,
$e_r(r)={n_0\phi_0\over c}v_\phi(r)$. 
The nonlocal Ohm's law takes the form
\begin{equation}
\xi_\perp^2{1\over r}{d\over dr}r{d\over dr}e_r(r)-\Big(1+{\xi_\perp^2\over r^2}\Big)e_r(r)=-{\alpha_L\over \gamma_\perp r}
\label{hydrophi}
\end{equation}
where 
$\alpha_L=n_0\phi_0I/(2\pi Wc)$ controls the strength of the Lorentz force.
The general solution of Eq. \ref{hydrophi} is given by
\begin{equation}
\label{radial_er}
e_r(r)={\rho_\perp I\over 2\pi W\xi_\perp}\Big[{\xi_\perp\over r}+c_1I_1(r/\xi_\perp)
   +c_2K_1(r/\xi_\perp)\Big],
\end{equation}
where $I_1$ and $K_1$ are Bessel functions.
The constants $c_1$ and $c_2$ are determined by 
requiring the liquid flow velocity, and therefore the electric field, to vanish 
at the boundaries with the Bose glass regions,
$e_r(R_1)=e_r(R_2)=0$, with the result,
\begin{eqnarray}
\label{constants}
& & c_1={K_1(\rho_2)/\rho_1-K_1(\rho_1)/\rho_2\over
        K_1(\rho_1)I_1(\rho_2)-K_1(\rho_2)I_1(\rho_1)} \\ \nonumber
& & c_2={I_1(\rho_1)/\rho_2-I_1(\rho_2)/\rho_1\over
        K_1(\rho_1)I_1(\rho_2)-K_1(\rho_2)I_1(\rho_1)},
\end{eqnarray}
where $\rho_{1,2}=R_{1,2}/\xi_\perp$. As in the channel problem, spatial inhomogeneities
are controlled by the in-plane correlation length $\xi_\perp$.
The field profile for a few values of $\xi_\perp$ in a micron-scale sample is shown in Fig.~5.

\vspace*{0.2in}
\epsfxsize=3.in
\epsfbox{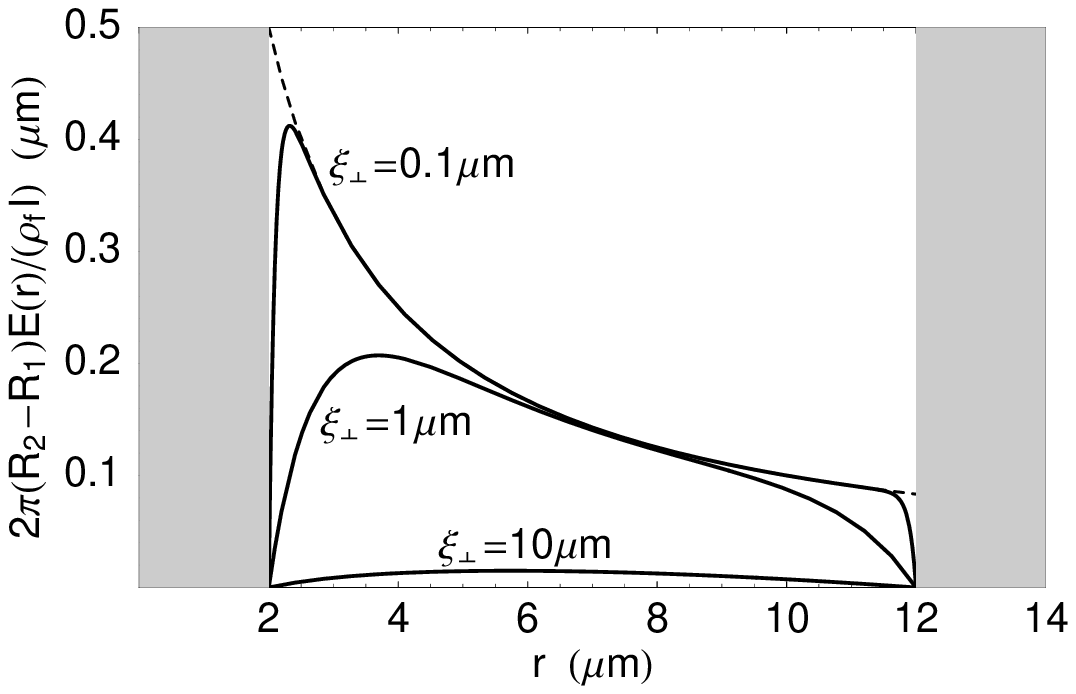}
\label{CorbinoProfiles}
\vspace{0.2in}

\noindent {\bf Figure 5}. The electric field profile for the Corbino 
disk geometry, with
$R_1=2\mu m$ and $R_2=12\mu m$, with an annular width $R_2-R_1=10\mu m$.
The vertical axis represents the reduced field $2\pi e_r(r)/(\rho_\perp I)$
in $\mu m$. 
The dashed line is the electric field for vanishing viscosity,
$e_r^0(r)={\rho_\perp I\over 2\pi(R_2-R_1)}{1\over r}$.
The shaded regions represent the Bose glass contacts.
\vspace{0.2in}

The field nonlocality arising from the growing correlation length near the Bose glass transition
can be probed experimentally by measuring the voltage profile across the annular channel.
A series of equally spaced contacts are placed radially, as shown in Fig.~5,
at positions $r_n=R_1+nd$, for $n=1,2,...,N$.
For the magnetic fields and temperatures of interest, the spacing $d$ 
of the contacts is large compared to the intervortex spacing $a_0$, so that hydrodynamics can be applied. 
The voltage drop 
across two successive contacts is obtained immediately by
integrating Eq. (\ref{radial_er}) and it is given by
\begin{eqnarray}
\Delta &&V(r_{n+1},r_n)= \int_{r_n}^{r_{n+1}} dr e_r(r)\\\nonumber
=&& {\rho_\perp I\over 2\pi W}\Big\{\ln(r_{n+1}/r_n)
    +c_1\big[I_0(r_{n+1}/\xi_\perp)-I_0(r_n/\xi_\perp)\big]\\\nonumber
 & &   -c_2\big[K_0(r_{n+1}/\xi_\perp)-K_0(r_n/\xi_\perp)\big]\Big\}.
\label{voltage}
\end{eqnarray}
The temperature dependence of the correlation length can then in principle be
extracted by fitting the measured voltage to Eq. (\ref{voltage}).

For comparison it is useful to discuss
the limiting cases of very small and very large correlation length.
Away from the Bose glass transition, where $\xi_\perp<<R_1,R_2,d$, viscous forces are negligible
and the boundary condition of vanishing flow velocity at the Bose-glass 
boundaries cannot be satisfied. The field from vortex motion is
simply 
\begin{equation}
\label{vel0}
e_r^0={\rho_\perp I\over 2\pi W}{1\over r}.
\end{equation}
The voltage drop across each pair of contacts is then given by
\begin{equation}
\label{voltage0}
\Delta V_0(r_{n+1},r_n)={\rho_\perp I\over 2\pi W}\ln(r_{n+1}/r_n)
\end{equation}
and is controlled by the bulk flux flow resistivity $\rho_\perp$ given in
Eq. (\ref{rhoperp2}). In this limit the vortex liquid only has very short ranged correlations;
the local electric field  is simply proportional to
the local current and decreases as $\sim 1/r$.
As a result, the voltage across every successive pair
of contacts decreases monotonically at all temperatures 
as one moves from the center to
the outer perimeter of the Corbino disk, i.e.,
$V_0(r_2,r_1)>V_0(r_3,r_2)>...>V_0(r_{n+1},r_n)$.

Near the transition, when the correlation length becomes comparable to the channel width,
viscous effects are dominant and the field profile can be approximated by
\begin{equation}
\label{er_infinity}
e_r^\infty(r)={\rho_\perp I\over 4\pi W\xi_\perp}\Big[b_1 r-{b_2\over r}-r\ln(r/\xi_\perp)\Big],
\end{equation} 
with $b_1,b_2>0$ given by
\begin{eqnarray}
& & b_1={R_2^2\ln(R_2/\xi_\perp)-R_1^2\ln(R_1/\xi_\perp)\over R_2^2-R_1^2},\\ \nonumber
& & b_2={R_1^2R_2^2\over R_2^2-R_1^2}\ln(R_2/R_1).
\end{eqnarray}
The voltage drop across two successive contacts is then given by
\begin{eqnarray}
\label{Vinfinity}
\Delta V_\infty(r_{n+1},r_n)={\rho_\perp I\over 4\pi W\xi_\perp}
  \bigg\{ & &{2b_1+1\over 4}\Big(r_{n+1}^2-r_n^2\Big) 
  -b_2\ln\big(r_{n+1}/r_n\big)\nonumber\\
& &  -{1\over 2}\Big[r_{n+1}^2\ln\big(r_{n+1}/\xi_\perp\big)
     -r_{n}^2\ln\big(r_{n}/\xi_\perp\big)\Big]\bigg\}.
\end{eqnarray}
The first term on the right hand side of Eq. \ref{er_infinity} 
can be easily understood
as it corresponds to the field
resulting from the rigid body rotation of a vortex solid at an angular
frequency $\omega=n_0\phi_0 I/(8\pi cW\eta)$. In this rigid body limit
the electric field grows with $r$. \cite{lopez_corbino} As a result, 
the voltage across two successive contacts increases at all temperatures 
as one moves from the center to the outer perimeter of the Corbino disk,
i.e., $V_0(r_2,r_1)<V_0(r_3,r_2)<...<V_0(r_{n+1},r_n)$.

In general as the temperature is decreased towards the Bose glass transition,
one will observe a crossover from liquid-like ($e_r\sim 1/r$)
to solid-like ($e_r\sim r$) response of the vortex array.

The voltages  across each successive pair of five contacts
are plotted in Fig.~6 as functions of $d/\xi_\perp(T)$. The correlation
length $\xi_\perp(T)$ increases with temperature as the Bose glass
transition is approached from above. The figure therefore displays
the growth of the voltage with increasing temperature. 
At high 
temperature, where $\xi_\perp<<d$, the voltage is given by 
Eq. (\ref{voltage0}), which is scaled out, and all curves approach 1.
The scaled voltage drop across successive pair of contacts is no
longer a monotonic
function of temperature. Rather it first increases as then decreases as one 
moves from the center towards the outer border of the disk.

\vspace*{0.2in}
\epsfxsize=3.in
\epsfbox{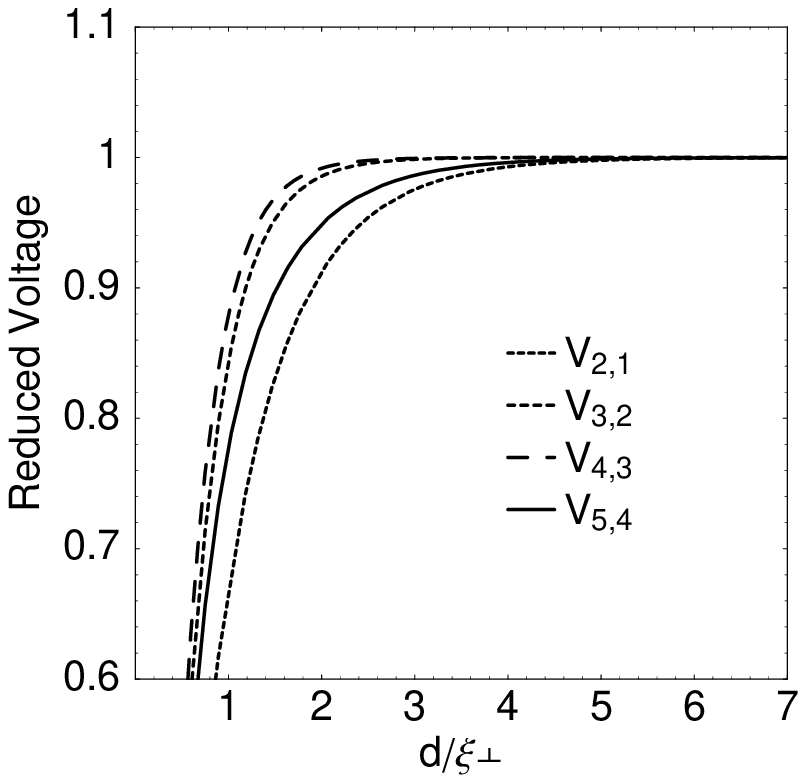}
\label{voltage_plots}
\vspace{0.2in}

\noindent {\bf Figure 6}. The voltage drop across a series of equally spaced
contacts located radially at $r_n=R_1+nd$, for $n=1,2,...,5$.
The vertical axis represents the voltage across two neighboring contacts
scaled by the corresponding voltage obtained for $\xi_\perp<<d$,
i.e.,
$V_{n+1,n}=\Delta V(r_{n+1},r_n)/\Delta V_0(r_{n+1},r_n)$,
where $\Delta V(r_{n+1},r_n)$ and $\Delta V_0(r_{n+1},r_n)$ are given
by Eqs. (4.13) and (4.15), respectively.
We have chosen $R_1=d$ and $R_2=6d$. The voltage is plotted as a function 
of $d/\xi_\perp$ which increases with temperature, as discussed in the text.
\vspace{0.2in}

\section{Response at finite frequency}
\label{sec:frequency}

The scaling theory of the continuous Bose glass transition in bulk samples 
is easily generalized to a finite frequency $\omega$, as described in 
Ref. \onlinecite{drnleo}. This is simply done by the addition of another 
scaling combination $\sim\omega l^z$ to the scaling functions of 
Eqs. (\ref{Eperp_bs}) and (\ref{Epar_bs}).
Similarly, we can generalize to finite frequencies the scaling theory  for 
spatially inhomogeneous flow in constrained geometries.
For simplicity we refer again to the channel geometry of Fig.~2 and only discuss the response
of the vortex liquid to a spatially homogeneous current of finite frequency $\omega$
applied across the channel in the $ab$ plane.
The generalized scaling ansatz based on a renormalization group flow for the spatially inhomogeneous field
from flux motion is,
\begin{equation}
\label{freq_scaling}
e_\perp(T,J_\perp,\omega,x,L)=l^{-(1+z)}e_\perp\big(l^{1/\nu}t, l^{\nu(1+\zeta)}J_\perp,l^z\omega\tau_0,l^{-1}x,l^{-1}L\big),
\end {equation}
with $\tau_0$ a microscopic time scale.\cite{tau0}
The flow diagram is sketched in Fig.~7. 
Clearly the behavior near the critical point depends on how the transition is approached in this plane.
For $\omega=0$ we 
fix the length $l$ as before by choosing $\tilde{t}(l_*)=t l_*^{1/\nu}=1$, which gives 
$\xi_\perp(T)\sim l_*=t^{-\nu}$ and yields the temperature dependence discussed earlier for
the dc response near the transition. The same analysis should hold at finite, but low frequencies,
provided at $l_*$ we have $\tilde{t}(l_*)>>\tilde{\omega}(l_*)$, with $\tilde{\omega}(l)=\omega\tau_0 l^z$.
This corresponds to the condition $\omega<<1/\tau(T)$, with $\tau(T)\sim\xi_\perp^z$
the relaxation time of critical fluctuations.
\begin{center}
\vspace{0.2in}
\epsfxsize=3.5in
\epsfbox{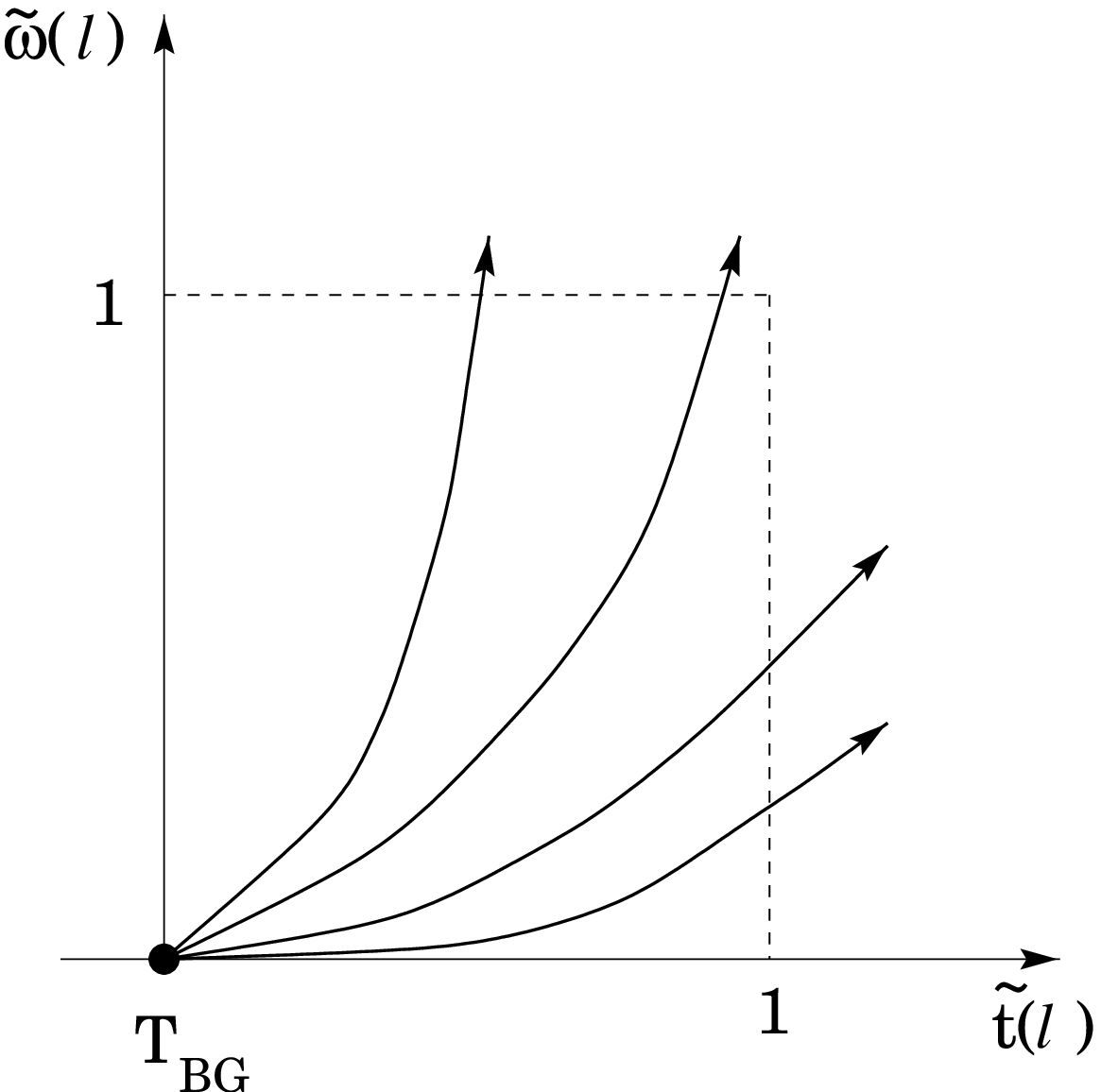}
\label{Rgflow}
\end{center}
\vspace{0.1in}
\noindent {\bf Figure 7}. A sketch of the flow diagram in the $(\tilde{\omega},\tilde{t})$ plane for a few initial
values of $\omega\tau_0/t^{\nu z}$.
\vspace{0.2in}

Conversely, we can approach the transition  along the frequency axis by letting $T=T_{BG}$ and using a 
finite frequency to tune the distance from the critical point. In this case we
fix $l$ by choosing $\tilde{\omega}(l_*)=1$, which gives 
$\xi_\perp^\omega\sim l_*=\omega^{-1/z}$. If the transition is approached along the
$T=T_{BG}$ line, the scaling ansatz becomes,
\begin{equation}
\label{freq_fin}
e_\perp(T=T_{BG},J_\perp,\omega,x,L)=(\xi_\perp^\omega)^{-(1+z)}{\cal E}_\omega\Big((\xi_\perp^\omega)^{\nu(1+\zeta)}J_\perp,
(\xi_\perp^\omega)^{-1}x,(\xi_\perp^\omega)^{-1}L\Big).
\end{equation}
As before, we are interested in the response in the vortex liquid phase, where 
the current-voltage characteristic is linear at small currents and we can write
\begin{equation}
\label{freq_lin}
e_\perp(T=T_{BG},J_\perp\rightarrow 0,\omega,x,L)=
  \rho_\perp^0\bigg({\xi_\perp^\omega\over a_0}\bigg)^{2-z}J_\perp{\cal F}_\omega\Big(
  x/\xi_\perp^\omega,L/\xi_\perp^\omega\Big).
\end{equation}
By integrating the field across the channel we immediately obtain the net voltage drop and
the corresponding ac resistivity as,
\begin{equation}
\label{rhoac}
\rho_{\perp L}(T_{BG},L,\omega)=\rho_\perp(T_{BG},\omega)f_\omega\Big(L/\xi_\perp^\omega\Big),
\end{equation}
where 
\begin{equation}
\label{rhoac_bulk}
\rho_{\perp}(T_{BG},\omega)=\rho_\perp^0\bigg({\xi_\perp^\omega\over a_0}\bigg)^{2-z}
  \sim (i\omega)^{1-2/z}.
\end{equation}
is the ac resistivity in bulk samples. \cite{drnleo} Both real and imaginary parts of
the linear frequency-dependent resistivity vanish as $\omega\rightarrow 0$ with the same
critical exponent, as obtained earlier in Ref. \onlinecite{drnleo}.
The scaling function in Eq. (\ref{freq_lin}) is simply
$f_\omega(x)={1\over x}\int_{-x/2}^{+x/2} du {\cal F}_\omega(u,x)$.
For $x>>1$ the ac resistivity in the channel must reduce to the bulk result,
which requires $f_\omega(x>>1)=1$.

To determine the form of the scaling function we assume again that at long 
wavelengths the local field from flux motion is described by the hydrodynamic 
equations discussed in section \ref{sec:hydro}.  The use of hydrodynamics for describing the ac 
response of current-driven vortex liquids has been discussed in Ref. \onlinecite{mcm_ac}.
Both friction and viscosity coefficients are in general frequency-dependent.
For the simple channel geometry, there is again only one hydrodynamic equation for 
the Fourier components of the the flow velocity at frequency $\omega$, given by
\begin{equation}
\label{ac_hyd}
-\gamma_\perp(\omega){\bf v}+\eta(\omega)\nabla^2{\bf v}={1\over c}n_0\phi_0\hat{\bf z}\times{\bf J}(\omega).
\end{equation}
Determining the frequency dependence of the friction and viscosity coefficients
 requires a microscopic calculation that is beyond the scope of the present paper.
However, their scaling with frequency at the transition can be readily obtained.
The hydrodynamic equation (\ref{ac_hyd}) defines a frequency dependent 
viscous length, $\sqrt{\eta(\omega)/\gamma_\perp(\omega)}$.
It is natural to identify this length with the diverging Bose glass correlation length,
\begin{equation}
\xi_\perp^\omega\sim\sqrt{{\eta(\omega)\over\gamma_\perp(\omega)}}.
\end{equation}
The complete scaling function defined by Eq. (\ref{freq_lin})
is then readily obtained by solving Eq. 
(\ref{ac_hyd}) for no slip boundary conditions and it is found to have the same functional form as that obtained for the dc response.
The ac resistivity for the channel geometry if given by
\begin{equation}
\rho_{\perp L}(T_{BG},L,\omega)=\rho_{\perp}(T_{BG},\omega)\bigg[1-{2\xi_\perp^\omega\over L}\tanh\bigg({L\over 2\xi_\perp^\omega}\bigg)\bigg],
\end{equation}
where $\rho_{\perp}(T_{BG},\omega)$ is the bulk ac resistivity given in 
Eq. (\ref{rhoac_bulk}). As for the dc case, one obtains the following behavior,
\begin{eqnarray}
& &\rho_{\perp L}(T_{BG},L,\omega)\approx\Big({n_0\phi_0\over c}\Big)^2{1\over 
\gamma_\perp(\omega)}\sim (i\omega)^{1-2/z}, \hskip 0.2in
    {\rm for}\hskip 0.2in \xi_\perp^\omega<<L,\\
& &\rho_{\perp L}(T_{BG},L,\omega)\approx \Big({n_0\phi_0\over c}\Big)^2{L^2\over 
  12\eta(\omega)}\sim L^2(\xi_\perp^\omega)^{-z}\sim L^2(i\omega),
  \hskip 0.2in{\rm for} \hskip 0.2in\xi_\perp^\omega<<L.
\end{eqnarray}
The condition $\xi_\perp^\omega\sim L$ defines a sample-dependent crossover frequency
$\omega_L=(a_0/L)^z/\tau_0\sim L^{-z}$ below which the channel geometry becomes important 
and $\rho_{\perp L}\sim i\omega L^2$. These results describe a Meissner-type response
of the superconductor that arises because the vortices are pinned and
immobile in the Bose glass phase.

The behavior observed as the critical point is approached along an arbitrary line in the $(\omega,t)$
plane depends on the location of that line. In general we expect that the $\omega=0$ scaling results
apply for $\omega<<1/\tau(T)\sim t^z$, while the scaling for $T=T_{BG}$ will apply for 
$\omega>>1/\tau$. The various regions for the expected scaling of the ac response are 
sketched in Fig.~8. Here the temperature scale $t_L$ is defined by $\xi_\perp(T)\sim L$,
with $t_L\sim (a_0/L)^{1/\nu}$. For $t>t_L$ the vortex array is insensitive to the confined 
geometry and the response is that of a bulk sample. For temperatures below $t_L$, the geometry becomes
important.
\begin{center}
\vspace{0.2in}
\epsfxsize=3.5in
\epsfbox{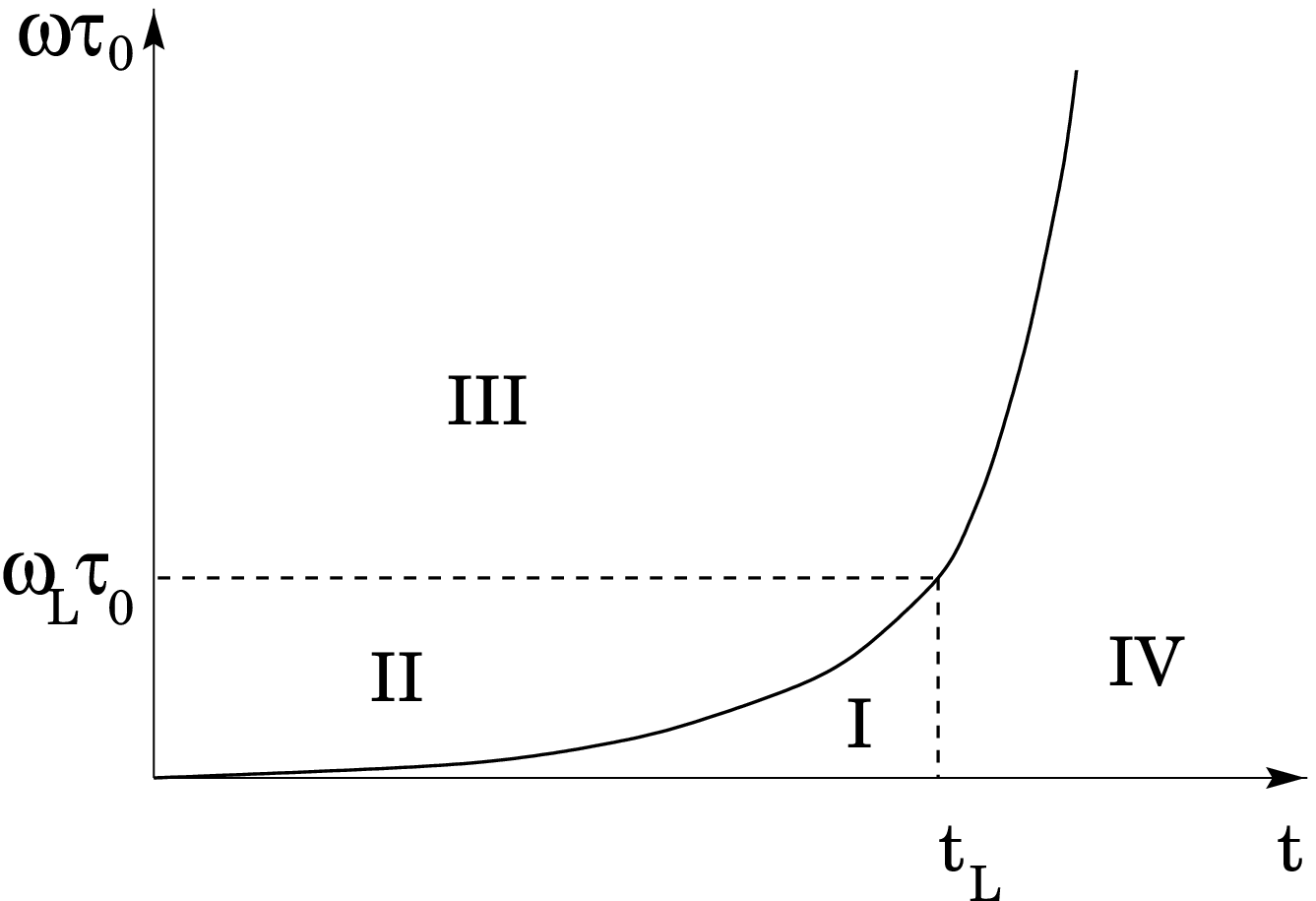}
\label{rhoac_diag}
\end{center}
\vspace{0.1in}
\noindent {\bf Figure 8}. A sketch of the scaling behavior expected for the ac response in 
various regions of the ($\omega,t)$ plane. The solid line is a plot of $\omega=1/\tau(T)=t^{z\nu}/\tau_0$.
The frequency and temperature scales $\omega_L$ and $t_L$ represent the scales below which the confined sample
geometry becomes important. They are defined by $\xi_\perp^{\omega_L}= L$, or
$\omega_L\tau_0=(a_0/L)^z$, and by $\xi_\perp(T_L)=L$, which gives $t_L=(a_0/L)^{1/\nu}$.
The ac resistivity $\rho_{\perp L}(\omega, T)$ is expected to scale as follows deep in each of the four regions sketched in the figure:
region I, $\rho_{\perp L}\sim L^2t^{\nu z}$;
region II, $\rho_{\perp L}\sim L^2(i\omega)$;
region III, $\rho_{\perp L}\sim (i\omega)^{1-2/z}$;
region IV, $\rho_{\perp L}\sim t^{\nu(z-2)}$.
\vspace{0.2in}

\vspace{0.4in}
This work was supported by the National Science Foundation at Syracuse through Grants No. DMR97-30678 and DMR98-05818
and at Harvard through Grant No. DMR97-14725, and by the Harvard Materials Research Science and Engineering Center
through Grant No. DMR98-09363.

\appendix
\section{Comparison with the Hydrodynamic Formulation of Mou {\it et al.}}

Mou et al. have used time-dependent Ginzburg-Landau theory to evaluate the nonlocal dc conductivity
tensor $\Sigma_{\mu\nu}({\bf r},{\bf r'})$ defined in Eq. (\ref{ohm}). Assuming nonlocal effects are slowly varying on the length scale
of interest, one can expand the right hand side of Eq. (\ref{ohm}) retaining terms up to quadratic 
in the gradients of the electric field. At long wavelengths, the nonlocal relation between fields and currents then takes
the form,
\begin{equation}
j_\mu({\bf r})=\sigma_{\mu\nu}({\bf 0})e_\nu({\bf r})-S_{\mu\alpha\beta\nu}\partial_\alpha\partial_\beta
   e_\nu({\bf r}),
\label{mou_cond}
\end{equation}
where $\sigma_{\mu\nu}({\bf 0})$ is the usual long-wavelength conductivity and the tensor $S_{\mu\alpha\beta\nu}$
describes the effect of nonlocality to quadratic order in the gradients. This fourth order tensor has six independent
components that were evaluated in Ref. \onlinecite{mou}. 

Upon comparing Eq. (\ref{mou_cond}) to Eq. (\ref{Qconstitutive}) and recalling the relationship between 
the fields and the vortex densities and fluxes given by the long-wavelength limit of
Eqs. (\ref{london_b}) and (\ref{london_eT}),
\begin{eqnarray}
& & {\bf b}=\phi_0{\bf T},\nonumber\\
& & e_\mu={\phi_0\over c}\epsilon_{\mu\nu\j}Q_{\nu j},
\end{eqnarray}
we see immediately that the local conductivity tensor  $\sigma_{\mu\nu}({\bf 0})$ and 
the tensor $S_{\mu\alpha\beta\nu}$ are simply proportional to the friction and viscosity tensor,
respectively,
according to,
\begin{eqnarray}
\label{gammatens}
& & \sigma_{\gamma\sigma}({\bf 0})={c^2\over \phi_0^2}\epsilon_{\gamma\mu\nu}
    \tilde{\gamma}_{\mu\nu,\lambda\rho}\epsilon_{\lambda\rho\sigma},\\
\label{etatens}
& & S_{\gamma\alpha\beta\sigma}={c^2\over \phi_0^2}
   \epsilon_{\gamma\mu\nu}\tilde{\eta}_{\mu\nu,\alpha\beta,\lambda\rho}
   \epsilon_{\lambda\rho\sigma}.
\end{eqnarray}

Neglecting Hall voltages, the local conductivity tensor is diagonal and has only two independent components, due
to the rotational symmetry in the plane normal to the applied field ($xy$ plane),
$\sigma_{xx}({\bf 0})=\sigma_{yy}({\bf 0})$ and $ \sigma_{zz}({\bf 0})$, which gives
\begin{equation}
\sigma_{\alpha\beta}({\bf 0})=\sigma_{xx}({\bf 0})\delta^\perp_{\alpha\beta}
             +\sigma_{zz}({\bf 0})\delta_{\alpha z}\delta_{\beta z},
\end{equation}
where 
$\delta^\perp_{\alpha\beta}=\delta_{\alpha_\perp,\beta_\perp}=\delta_{\alpha\beta}-\delta_{\alpha z}\delta_{\beta z}$
and the labels with a $\perp$ subscript, e.g., $\alpha_\perp,\beta_\perp,...$, only run over the two values $x$ and $y$.
The friction tensor is given by
\begin{equation}
\tilde{\gamma}_{\mu\nu,\lambda\rho}={1\over 2n_0^2}\Big(\gamma_\perp\epsilon_{\mu\nu\alpha_\perp}\epsilon_{\alpha_\perp\lambda\rho}
   +\gamma_\parallel\epsilon_{\mu\nu z}\epsilon_{\alpha\lambda z}\Big),
\end{equation}
with
\begin{eqnarray}
& & \sigma_{xx}({\bf 0})=\Big({c\over n_0\phi_0}\Big)^2\gamma_\perp,\nonumber\\
& & \sigma_{zz}({\bf 0})=\Big({c\over n_0\phi_0}\Big)^2\gamma_\parallel.
\end{eqnarray}
Symmetry under rotations about the $z$ direction also restricts the number of independent components of the 
viscosity tensor and the related tensor
$S_{\mu\alpha\beta\nu}$  to six.
The tensor $S_{\mu\alpha\beta\nu}$ is given by
\begin{eqnarray}
S_{\mu\alpha\beta\nu}=& & S_{xxxx}\delta^\perp_{\mu\nu}\delta^\perp_{\alpha\beta}
   +\big(S_{xyyx}-S_{xxxx}\big)\Big[\delta^\perp_{\mu\nu}\delta^\perp_{\alpha\beta}
             -{1\over 2}\big(\delta^\perp_{\mu\alpha}\delta^\perp_{\beta\nu}+
                            \delta^\perp_{\mu\beta}\delta^\perp_{\alpha\nu}\big)\Big]\nonumber \\
& &  +S_{xzzx}\delta^\perp_{\mu\nu}\delta^{\alpha z}\delta^{\beta z}
     +S_{zxxz}\delta^\perp_{\alpha\beta}\delta^{\mu z}\delta^{\nu z}
     +S_{zzzz}\delta_{\mu z}\delta_{\alpha z}\delta_{\beta z}\delta_{\nu z}\nonumber\\
& &  +S_{xxzz}\big(\delta_{\alpha z}\delta_{\mu z}\delta^\perp_{\beta\nu}
                 +\delta_{\beta z}\delta_{\nu z}\delta^\perp_{\alpha\mu}
                 +\delta_{\beta z}\delta_{\mu z}\delta^\perp_{\alpha\nu}
                 +\delta_{\alpha z}\delta_{\nu z}\delta^\perp_{\beta\mu}\big).
\end{eqnarray}
The viscosity tensor is given by
\begin{eqnarray}
\tilde{\eta}_{\mu\nu,\alpha\beta,\lambda\rho}={1\over 2n_0^2}\bigg\{ & &
       \eta\delta^\perp_{\alpha\beta}\epsilon_{\mu\nu\gamma_\perp}\epsilon_{\gamma_\perp\lambda\rho}
      +\eta_b\Big[\delta^\perp_{\alpha\beta}\epsilon_{\mu\nu\gamma_\perp}\epsilon_{\gamma_\perp\lambda\rho}
           -{1\over 2}\big(\epsilon_{\mu\nu\alpha_\perp}\epsilon_{\beta_\perp\lambda\rho}
                       +\epsilon_{\mu\nu\beta_\perp}\epsilon_{\alpha_\perp\lambda\rho}\big)\Big]\nonumber\\
& &   +\eta_z\delta_{\alpha z}\delta_{\beta z}\epsilon_{\mu\nu\gamma_\perp}\epsilon_{\gamma_\perp\lambda\rho}
      +\eta^t \delta^\perp_{\alpha\beta}\epsilon_{\mu\nu z}\epsilon_{z\lambda\rho}
      +\eta^t_z\delta_{\alpha z}\delta_{\beta z}\epsilon_{\mu\nu z}\epsilon_{z\lambda\rho}\nonumber\\
& &   +{1\over 2}\eta_x\Big[\delta_{\alpha z}\big(\epsilon_{\mu\nu z}\epsilon_{\beta_\perp\lambda\rho}
                                           + \epsilon_{\mu\nu\beta_\perp}\epsilon_{z\lambda\rho}\big)
                            +\delta_{\beta z}\big(\epsilon_{\mu\nu z}\epsilon_{\alpha_\perp\lambda\rho}
                                           + \epsilon_{\mu\nu\alpha_\perp}\epsilon_{z\lambda\rho}\big)\Big]
\bigg\},
\end{eqnarray}
with
\begin{eqnarray}
\label{viscosities}
& & S_{xxxx}=\Big({c\over n_0\phi_0}\Big)^2 \eta,\nonumber\\
& & S_{xzzx}=\Big({c\over n_0\phi_0}\Big)^2 \eta_z,\nonumber\\
& & \big(S_{xyyx}-S_{xxxx}\big)=\Big({c\over n_0\phi_0}\Big)^2 \eta_b,\nonumber\\
& & S_{zxxz}=\Big({c\over n_0\phi_0}\Big)^2 \eta^t,\nonumber\\
& & S_{zzzz}=\Big({c\over n_0\phi_0}\Big)^2 \eta^t_z,\nonumber\\
& & S_{xxzz}={1\over 2}\Big({c\over n_0\phi_0}\Big)^2 \eta_x.
\end{eqnarray}

Mou et al. evaluated the lowest order fluctuation contribution to the
components of the tensor $S_{\mu\alpha\beta\nu}$ for a vortex liquid 
using a time-dependent Ginzburg-Landau theory within a Gaussian approximation.

\section{DC Flux Transformer Experiments}

DC flux transformer experiments have been used as probes of the nonlocal resistive response of 
vortex liquids. In particular, by comparing transport measurements in twinned and untwinned
$YBCO$, Lop\'ez et al. \cite{lopez_twinned} argued that in the twinned crystal, where the liquid transforms into a solid via
a continuous glass transition, measurements in the transformer configuration may provide a direct probe
of the longitudinal correlation length near the transition. Here we reconsider the 
analysis of such experiments in the context of hydrodynamics following earlier work by
Huse and Majumdar. \cite{huse}. Our objective is to make the connection between the 
measured voltages and the diverging correlation lengths
more precise. 
As pointed out by Huse and Majumdar, explicit calculations
of the current and voltage patterns have to be carried out for each specific sample
geometry by solving a fourth order partial differential equation with rather complicated boundary conditions.
Here we reformulate the problem discussed in Ref. \onlinecite{huse} incorporating 
all relevant viscosity coefficients, then discuss the solution in terms of diverging length scales. 
We also correct Eq. (7) of Ref. \onlinecite{huse}
where some boundary terms were left out.

We are interested in discussing dc transport experiments in finite materials of specified geometry, where 
the current is injected and withdrawn via contacts placed at the materials' boundaries. 
As discussed by Huse and Majumdar, in the bulk of the sample
in steady state the total current satisfies $\bbox{\nabla}\cdot{\bf j}=0$
and the electric field is irrotational, $\bbox{\nabla}\times{\bf e}=0$. 
The relationship between current and field is given by the nonlocal Ohm's law, Eq. (\ref{ohm}), which at long wavelength takes the form given in Eq. (\ref{mou_cond}). 
The field ${\bf e}$ can be written in terms of a local voltage as 
\begin{equation}
\label{fluxV}
{\bf e}=-\bbox{\nabla}\Phi,
\end{equation}
and the nonlocal Ohm's law transforms into a fourth order differential equation for the voltage, $\Phi$,
\begin{equation}
\label{bulk_voltage}
\Big[\sigma_{\mu\nu}({\bf 0})\partial_\mu\partial_\nu-S_{\mu\alpha\beta\nu}\partial_\mu\partial_\alpha
\partial_\beta\partial_\nu\Big]\Phi=0.
\end{equation}
The sample boundaries separate a region where the local field or vortex flow velocity can have spatial gradients
due to viscous forces among the vortices from the outside region where no such spatial gradients occur.
Neglecting nonlocalities on length scales of the order of the penetration length, the first spatial derivatives
of the field ${\bf e}$ (or equivalently of the vortex flux tensor $Q_{\mu\nu}$) can therefore exhibit 
jump discontinuities at the sample boundaries. Such discontinuities correspond to $\delta$-function
contributions to the local current, denoted by ${\bf j}^\delta$. 
Considering a local coordinate system with an $s$ axis normal to the sample boundary, located at $s=0$,
and directed inwards, the nonlocal Ohm's law at the boundary takes the form,
\begin{equation}
j_\mu({\bf r})+j^\delta_\mu({\bf r})=
   \sigma_{\mu\nu}({\bf 0})e_\nu({\bf r})-S_{\mu\alpha\beta\nu}\partial_\alpha\partial_\beta
   e_\nu({\bf r}) -\delta(s)S_{\mu ss\nu}\partial_\nu \Big[e_\nu({\bf r})\Big]_{s\rightarrow 0^+},
\label{mou_cond_boundaries}
\end{equation}
where the subscript $s$ is not summed over. The last term on the right hand side of the equation
equals the $\delta$ function contribution to the surface current. Such contribution must be flowing 
within the surface
of the sample, i.e., its component normal to the boundary must vanish,
\begin{equation}
\label{boundary_one}
j^\delta_s=S_{sss\nu}\Big[\partial_s e_\nu\Big]_{s\rightarrow 0^+}=
   -S_{sss\nu}\Big[\partial_s\partial_\nu\Phi\Big]_{s\rightarrow 0^+}=0.
\end{equation}
This provides the first set of boundary conditions for Eq. (\ref{bulk_voltage})
The second boundary condition is obtained by requiring that the $\delta$-function part of
the divergence of the current at the surface equals the total current $I_s(s)$ injected at that point of the surface,
\begin{equation}
\label{surface-div}
\big(\bbox{\nabla}\cdot{\bf j}\big)^\delta=\bbox{\nabla}\cdot{\bf j}^\delta=
-\delta(s)S_{\mu_\perp ss\nu}\partial_{\mu_\perp}\partial_\nu \Big[e_\nu({\bf r})
\Big]_{s\rightarrow 0^+}=I_s(s),
\end{equation}
where again the index $s$ is not summed over and $\mu_\perp$ only runs over the two components 
perpendicular to the surface.
This requires that the voltage at the surface satisfies,
\begin{equation}
\label{boundary_V}
\Big[\sigma_{s\nu}({\bf 0})\partial_\nu\Phi-S_{s\alpha\beta\nu}\partial_\mu\partial_\alpha
\partial_\beta\partial_\nu\Phi -S_{\mu_\perp ss\nu}\partial_{\mu_\perp}\partial_s\partial_\nu\Phi
-2S_{\mu_\perp\nu ss}\partial_{\mu_\perp}\partial_s\partial_\nu\Phi\Big]_{s=0^+}=I_s(s).
\end{equation}
The last term in square brackets on the left hand side of Eq. (\ref{boundary_V}) was absent in Ref. \onlinecite{huse}.

The specific transformer configurations of interest are shown in Fig.~9. We neglect spatial inhomogeneities 
in the $x$ direction (parallel to the flow) and reduce the problem to a two-dimensional one. The sample has finite
extent $W$ in the $y$ direction and thickness $L$ in the external field direction.

\vspace*{0.2in}
\epsfxsize=3.0in
\epsfbox{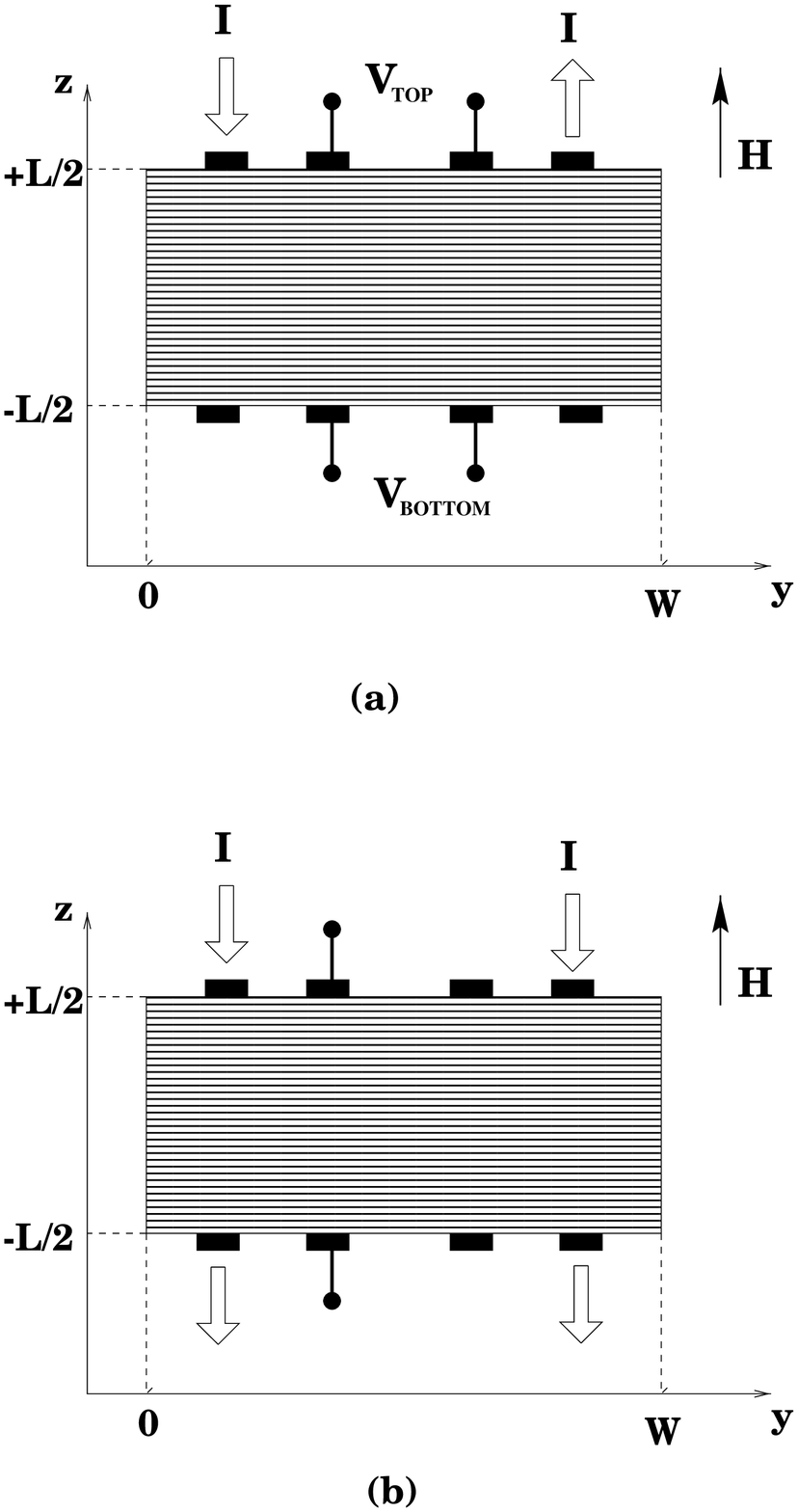}
\label{transformer}
\vspace{0.2in}

\noindent {\bf Figure 9}. The dc flux transformer geometry.
\vspace{0.3in}

Current is injected and withdrawn at the contacts shown placed on the upper and lower surface of the sample.
Therefore no current is injected at the boundaries normal to the $y$ direction.
In order to discuss the role of the various viscosity coefficients in controlling the 
nonlocal response of the vortex array, it is useful to return to our hydrodynamic 
equations in terms of vortex densities and fluxes. In the geometry of interest there are
two nonvanishing components of the electric field from flux motion, 
related to the corresponding
components of the vortex flux tensor as
\begin{eqnarray}
& &e_y(y,z)= {n_0\phi_0\over c} v_x(y,z),\nonumber \\
& &e_z(y,z)= {n_0\phi_0\over c} V(y,z).
\end{eqnarray}
From Eq. (\ref{Tcontinuity}) in a steady state vortex current is conserved,
which requires
\begin{equation}
\partial_zV=\partial_yv_x,
\end{equation}
This condition is equivalent to the requirement that 
$\bbox{\nabla}\times{\bf e}=0$ and implies that the components of the vortex flux tensor 
can be derived by a potential, $\Phi_v$, according to
\begin{eqnarray}
& &v_x(y,z)=\partial_y\Phi_v,\nonumber \\
& &V(y,z)=\partial_z\Phi_v,
\end{eqnarray}
where $\Phi_v$ is simply proportional to the voltage defined in Eq. (\ref{fluxV}),
$\Phi_v=-{c\over n_0\phi_0}\Phi$.
The hydrodynamic equations governing vortex flow in the transformer geometry shown
in Fig.~9 are
\begin{eqnarray}
\label{transf_vx}
& &\gamma_\perp v_x=\Big(\eta\partial_y^2\eta_z\partial_z^2\Big)v_x+\eta_x\partial_y
\partial_z V +{c_{44}\over n_0}\partial_zt_x+{n_0\phi_0\over c}J_y,\\
\label{transf_V}
& & \gamma_\parallel V=\Big(\eta^t\partial_y^2+\eta_z^t\partial_z^2\Big)V
+\eta+x\partial_y\partial_zv_x-{c_{44}\over n_0}\partial_yt_x
 +{n_0\phi_0\over c}J_z.
\end{eqnarray}
By differentiating Eq. (\ref{transf_vx}) with respect to $y$ and Eq. (\ref{transf_V})
with respect to $z$ and making use again of the fact that the total current 
is divergence-free in bulk, we immediately
obtain the equation for the potential $\Phi_v$,
\begin{equation}
\label{voltage4}
\gamma_\perp\partial_y^2\Phi_v+\gamma_\parallel\partial_z^2\Phi_v
-\Big[\eta\partial_y^4+(\eta_z+\eta^t+2\eta_x)\partial_y^2\partial_z^2
   +\eta_z^t\partial_z^4\Big]\Phi_v=0.
\end{equation}
This fourth order differential equation has to be solved with the boundary conditions,
\begin{eqnarray}
& & \eta\Big[\partial_y^2\Phi_v\Big]_{y=0,W}=0,\nonumber\\
& & \eta_z^t\Big[\partial_z^2\Phi_v\Big]_{z=0,L}=0,\nonumber\\
& & \Big[\gamma_\perp\partial_y\Phi_v-\eta\partial_y^3\Phi_v-(\eta_z+2\eta_x+\eta^t)
   \partial_z^2\partial_y\Phi_v\Big]_{y=0,W}=0,\nonumber\\
& & \Big[\gamma_\parallel\partial_z\Phi_v-\eta_z^t\partial_z^3\Phi_v-(\eta_z+2\eta_x+\eta^t)
   \partial_y^2\partial_z\Phi_v\Big]_{z=0,L}=\pm {c\over n_0\phi_0}I_z\big(y,z=\pm{L\over 2}\big).
\end{eqnarray}

To proceed, we follow Huse and Majumdar and neglect viscous drag associated with $\eta$ and $\eta_z^t$
 (the bulk viscosity $\eta_b$ drops out of the problem for linear incompressible flow).
This amounts to dropping the terms $\sim\partial_y^4$
and $\sim\partial_z^4$ compared to the fourth order cross derivatives in Eq. (\ref{voltage4}).
Letting $\tilde{\eta}_z=\eta_z+\eta^t+2\eta_x$ \cite{huse_eta}, the equation simply reduces to
\begin{equation}
\gamma_\perp\partial_y^2\Phi_v+\gamma_\parallel\partial_z^2\Phi_v
-\tilde{\eta}_z\partial_y^2\partial_z^2\Phi_v=0,
\end{equation}
with the boundary conditions
\begin{eqnarray}
& & \Big[\gamma_\perp\partial_y\Phi_v-\tilde{\eta}_z
   \partial_z^2\partial_y\Phi_v\Big]_{y=0,W}=0,\nonumber\\
& & \Big[\gamma_\parallel\partial_z\Phi_v-\tilde{\eta}_z
   \partial_y^2\partial_z\Phi_v\Big]_{z=0,L}=\pm {c\over n_0\phi_0}I_z(y,\pm L/2).
\end{eqnarray}

The solution for an arbitrary distribution of current sources and sinks has been given by
Huse and Majumdar. \cite{huse} Simply translating their results into our notation, 
we write the current sources at the top and bottom of the sample as
\begin{equation}
I(y,z=\pm L/2)=\sum_{n=0}^\infty i_n^\pm\cos(n\pi y/W).
\end{equation}
The integral of the current injected or withdrawn over the entire sample surface
must vanish. This requires $i_0^+=-i_0^-$.
Making use of the boundary conditions at $x=0,W$, the voltage is then written as
\begin{equation}
\Phi(y,z)=\sum_{n=0}^\infty U_n(z)\cos(n\pi y/W),
\end{equation}
where
\begin{eqnarray}
& & U_0(z)=C_0+S_0z,\nonumber\\
& & U_n(z)=C_n\cosh(K_nz)+S_n\sinh(K_nz), \hskip 0.2in  n>0,
\end{eqnarray}
with 
\begin{equation}
\label{screening}
K_n={n\pi\over W}\bigg({\gamma_\perp(T)\over\gamma_\parallel(T)}\bigg)^{1/2}
  \bigg[{1+{\tilde{\eta}_z(T)\over \gamma_\parallel(T)}
  \big(n\pi/W\big)^2}\bigg]^{-1/2}.
\end{equation}
The coefficient $C_0$ sets an arbitrary zero for the voltage. We will let $C_0=0$ below. 
The other 
coefficients are given by 
\begin{eqnarray}
& & S_0=\rho_\parallel i_0^-,\nonumber\\
& & S_n={\rho_\perp(i_n^--i_n^+)\over 2}{K_n (W/n\pi)^2\over\cosh(K_nL/2)},\hskip 0.2in  n>0,\nonumber\\
& & C_n=-{\rho_\perp(i_n^-+i_n^+)\over 2}{K_n(W/n\pi)^2\over \sinh(K_nL/2)}, \hskip 0.2in  n>0.
\end{eqnarray}

Now we use this solution to discuss the experimental finding of Lope\'z et al.
\cite{lopez_twinned} in twinned and untwinned samples of $YBCO$. 
These authors have carried out two sets of measurements. The first consists in
injecting current at the left contact at the top of the sample and extracting it at the 
right contact, as shown in Fig.~9. The voltages $V_{\rm TOP}$ and
$V_{\rm BOTTOM}$ across the two central contacts at the top and bottom of the sample
are then measured as functions of temperature. For this choice of current sources,
$i_n^-=0$, for all $n$,  and $i_0^+=0$.
The voltage at the top and bottom of the sample is then given by
\begin{equation}
\Phi(y,\pm L/2)=-{1\over 2}\rho_\perp\sum_{n=1}^\infty i_n^+\cos(n\pi y/W)
   K_n\Big({W\over n\pi}\Big)^2 \bigg[{\cosh(K_nL/2)\over\sinh(K_nL/2)}
       \pm{\sinh(K_nL/2)\over\cosh(K_nL/2)}\bigg].
\end{equation}
The temperature dependence of the voltages is controlled by that of the inverse
length $K_n$, given in Eq. (\ref{screening}), which in turn depends on the friction 
and viscosity coefficients which
diverge near the vortex glass and the
Bose glass transitions with the critical exponents discussed earlier. In particular we recall
that
\begin{eqnarray}
\label{gamma_ratio}
& &{\gamma_\perp(T)\over \gamma_\parallel(T)}={\gamma_\perp^0\over \gamma_\parallel^0}
\Big({\xi_\perp\over\xi_\parallel}\Big)^2\sim t^{2\nu(\zeta-1)},\\
\label{eta_ratio}
& & {\tilde{\eta}_z(T)\over\gamma_\parallel(T)}={\tilde{\eta}_z^0\over\gamma_\parallel^0}
\Big({\xi_\perp\over a_0}\Big)^2\sim t^{2\nu},
\label{def_xipar}
\end{eqnarray}
where we have extracted
the divergence by rewriting all transport properties in terms of their 
bare values (denoted by a superscript ``$^0$''), which 
depend only weakly on temperature at the transition.
The longitudinal correlation length will be defined as
\begin{equation}
\xi_\parallel^2=\tilde{\eta}_z(T)/\gamma_\perp(T),
\end{equation}

We begin by discussing the case of vanishing viscosity. Then $K_n$ is simply given by 
\begin{equation}
\label{Kn0}
K_n\approx K_n^0={n\pi\over W}\sqrt{\gamma_\perp\over\gamma_\parallel}
   ={n\pi\over W}\sqrt{\rho_\parallel\over\rho_\perp}\sim t^{2\nu(\zeta-1)}.
\end{equation}
It vanishes therefore at the Bose glass transition ($\zeta=2$), but it remains finite at the
isotropic vortex glass transition ($\zeta=1$).
Choosing for simplicity a current distribution with $i_1^+\not=0$ and $i_n^+=0$ for all
other values of $n$, the voltage for $\tilde{\eta}_z=0$ is given by
\begin{equation}
\Phi_0(y,\pm L/2)=-{1\over 2}\sqrt{\rho_\parallel\rho_\perp}i_1^+\cos(\pi y/W)\Big({W\over \pi}\Big)^2
     \bigg[{\cosh(K_1^0L/2)\over\sinh(K_1^0L/2)}
       \pm{\sinh(K_1^0L/2)\over\cosh(K_1^0L/2)}\bigg],
\end{equation}
with $K_1^0$ given by Eq. (\ref{Kn0}) for $n=1$.
Close to the glass transition where $K_1^0 \sim t^{2\nu(\zeta-1)}$ we can use $K_1^0L<<1$
to obtain,
\begin{equation}
\Phi_0(y,\pm L/2)\approx -\rho_\perp{W^2\over \pi^2 L} i_1^+\cos(\pi y/W),
\end{equation}
and the voltage at the top and bottom of the sample are equal.
Conversely, if $K_1^0L>>1$, we find
\begin{eqnarray}
& & \Phi_0(y,+L/2)\approx -\sqrt{\rho_\parallel\rho_\perp}{W\over \pi}i_1^+\cos(\pi y/W),\nonumber\\
& & \Phi_0(y,-L/2)\approx 0,
\end{eqnarray}
with corrections to the above expressions decreasing exponentially as
$\sim e^{-K_1^0L}$. In this case the vortices behave like uncorrelated pancakes
and the bottom voltage is much smaller than the one at the top.
In other words, there is a crossover at $K_1^0L\sim 1$,  corresponding to
$\rho_\parallel L^2/4\sim \rho_\perp W^2/\pi^2$, from a regime where the 
top voltage exceeds the bottom voltage to one where both voltages are the same.
Near the Bose glass transition, where $K_1^0\sim t^{2\nu}$, this crossover is sharp
and occurs at a temperature that depends on both the width $W$ and the thickness $L$of the sample
and is given by $t_*^\nu\sim (2W/\pi L)\sqrt{\gamma_\parallel^0/\gamma_\perp^0}$.
In contrast, near the vortex glass transition where $\zeta=1$,
$K_1^0$ is weakly temperature dependent and no sharp crossover is expected.

Consider now the response in the case where the viscosity $\tilde{\eta}_z$ is very large,
with $(\tilde{\eta}_z/\gamma_\parallel)(\pi/W)^2>>1$, so that for all values of $n$,
\begin{equation}
K_n\approx K_n^\infty= \sqrt{\gamma_\perp\over\tilde{\eta}_z}={1\over\xi_\parallel}.
\end{equation}
In this case the top and bottom voltages for the first experiment (Fig.~9) are given by
\begin{equation}
\Phi_\infty(y,\pm L/2)\approx -{1\over 2}{\rho_\perp\over\xi_\parallel}
   \bigg[{\cosh(L/2\xi_\parallel)\over\sinh(L/2\xi_\parallel)}
       \pm{\sinh(L/2\xi_\parallel)\over\cosh(L/2\xi_\parallel)}\bigg]
      \sum_{n=1}^\infty i_n^+\cos(n\pi y/W)
   \Big({W\over n\pi}\Big)^2 .
\end{equation}
The crossover is now controlled by the length $\xi_\parallel\sim t^{-\zeta\nu}$.
If $L/2\xi_\parallel>>1$, the voltage reduces to
\begin{eqnarray}
& &\Phi_\infty(y,+ L/2)\approx -{\rho_\perp\over\xi_\parallel}
      \sum_{n=1}^\infty i_n^+\cos(n\pi y/W)
   \Big({W\over n\pi}\Big)^2,\nonumber\\
& &\Phi_\infty(y,- L/2)\approx 0,
\end{eqnarray}
i.e., voltages are uncorrelated well above the glass transition. 
As the glass transition is approached and $L/2\xi_\parallel <<1$, we obtain,
\begin{equation}
\Phi_\infty(y,\pm L/2)\approx -{\rho_\perp\over L}
      \sum_{n=1}^\infty i_n^+\cos(n\pi y/W)
   \Big({W\over n\pi}\Big)^2 .
\end{equation}
In this case the top and bottom voltages are the same and they are both reduced by
a factor $\xi_\parallel/L$ compared to the top voltage at high temperature.
Their vanishing at the transition is controlled by the vanishing of $\rho_\perp$.
We find a sharp crossover from a high temperature region where the top voltage exceeds
the bottom voltage to a low temperature region where the two voltages are equal.
the crossover temperature $T^*$ depends on the sample thickness and is given by 
$\xi_\parallel (T^*)\sim L/2$. It is therefore expected to decrease with sample
thickness according to $T^*\sim T_{BG}(L/a_0)^{-1/\zeta\nu}$, consistent
with the observations of Lope\'z et al. in twinned samples. 
For $T<T^*$ the voltages are the same and decreasing with decreasing temperature,
vanishing at $T_{BG}$ as $\rho_\perp\sim t^{\nu(z-\zeta)}$.

The second set of measurements carried out by Lope\'z et al. consists
of injecting a current at the top surface and extracting it at the bottom surface,
while measuring the voltage across two contacts at $z=\pm L/2$ and at the same $y$
position, as sketched in Fig.~9 .
The corresponding distribution of current sources and sinks corresponds to
$i_n^+=-i_n^-$, for every $n$. The voltage difference between top and bottom
is given by
\begin{eqnarray}
\Delta\Phi(y)& \equiv& \Phi(y,L/2)- \Phi(y,-L/2)\nonumber\\
&=& i_0^-\rho_\parallel L+2\rho_\perp\sum_{n=1}^\infty i_n^-\cos(n\pi y/W)
   K_n\Big({W\over n\pi}\Big)^2\tanh(K_nL/2).
\end{eqnarray}
For simplicity we assume the currents injected at the top and withdrawn at the bottom are
uniform in $y$, i.e., $i_0^-=-i_0^+={\rm constant}$, and $i_n^-=-i_n^+=0$ for $n\not=1$.
the voltage is then simply
\begin{equation}
\Delta\Phi(y)=i_0^-\rho_\parallel L
\end{equation}
and for all values of $y$ it is controlled by the $c$-axis resistivity $\rho_\parallel$,
which vanishes at the glass transition.


%


\begin{references}
\bibitem{tau0}
Here $\tau_0$ represents the relaxation time of a fluctuation of volume $\sim a_0^2\xi_z$,
with $\xi_z=(\tilde{\epsilon}_1/k_BT)a_0^2$ the characteristic spacing between entanglements
introduced in Ref. [4]. As shown in Ref. [24b], it can be related to the normal state resistivity.
It is the time necessary for a segment of size 
$\xi_z$ to diffuse an intervortex spacing, $\tau_0=a_0^2/D$, with $D$ the diffusion 
constant of a segment of length $\xi_z$. This is in turn given by
$D=D_0(c_0/\xi_z)$, with $D_0$ the diffusion constant of a vortex in an isolated copper-oxide plane, and $c_0$ the average spacing between $CuO_2$ planes. As shown in [24b],
one finds $\tau_0\approx (a_0\tilde{\epsilon}_1/k_BT)^2c^2/(BH_{c2}c_0\sigma_n^{2d})$,
with  $\sigma_n^{2d}$ the two-dimensional normal-state conductivity of an
isolated $CuO_2$ layer.
\bibitem{huse_eta}
In Ref. \onlinecite{huse} only the viscosity coefficient
$\eta_z$ was retained from the outset in the hydrodynamic equations.
\end{references}
\end{document}